\documentclass[11pt]{article}
\usepackage[T1]{fontenc}
\usepackage[utf8]{inputenc}
\usepackage{newpxtext}  
\usepackage{bm,amsmath,amssymb,mathtools,graphicx}
\usepackage{textcomp,gensymb,units,upquote}
\usepackage[a4paper]{geometry}
\usepackage{epsfig}
\usepackage[parfill]{parskip}
\usepackage[matrix,arrow,color]{xy}
\usepackage[usenames]{color}
\usepackage{mathrsfs}
\usepackage[font={small}]{caption}
\usepackage{subcaption}
\usepackage[export]{adjustbox}
\usepackage{float}
\usepackage{wrapfig}
\usepackage{microtype}
\usepackage{slashed}
\usepackage{braket}
\usepackage[pdftex]{hyperref}

\input{xy}
\xyoption{all}

\input{epsf}

\makeatletter

\usepackage{setspace}

\usepackage{lscape}

\catcode`@=11
\renewcommand{\section}
{\@startsection{section}{1}{0pt}{\medskipamount}{\medskipamount}{\large\bf}}
\makeatletter\renewcommand{\subsection}
{\@startsection{subsection}{2}{\z@}{-3.25ex plus -1ex minus -.2ex}
{1.5ex plus .2ex}{\it }}

\numberwithin{equation}{section}
\catcode`@=12

\newcommand{\ban}{\begin{eqnarray}}
\newcommand{\ean}{\end{eqnarray}}

\newcommand{\Tr}{{\rm Tr}}
\newcommand{\tr}{{\rm tr}}

\newcommand{\cW}{{\cal W}}
\newcommand{\cN}{{\cal N}}

\newcommand{\cB}{{\cal B}}

\newcommand{\cH}{{\cal H}}
\newcommand{\cA}{{\cal A}}
\newcommand{\cE}{{\cal E}}
\newcommand{\cO}{{\cal O}}
\newcommand{\cC}{{\cal C}}

\newcommand{\cR}{{\cal R}}

\newcommand{\cF}{{\cal F}}

\newcommand{\cK}{{\cal K}}

\newcommand{\cV}{{\cal V}}

\newcommand{\sfa}{{\mathsf{a}}}
\newcommand{\sfb}{{\mathsf{b}}}

\newcommand{\sfC}{{\mathsf{C}}}

\newcommand{\sfv}{{\mathsf{v}}}
\newcommand{\sfw}{{\mathsf{w}}}

\newcommand{\scrC}{{\mathscr{C}}}

\newcommand{\complex}{{\mathbb C}} 
\newcommand{\zed}{{\mathbb Z}} 
\newcommand{\real}{{\mathbb R}} 

\def\e{{\,\rm e}\,}

\def\ii{{\,{\rm i}\,}}
\def\dd{{\rm d}}

\newcommand{\sign}{\mathrm{sgn}}

\def\beq{\begin{equation}}
\def\bee{\begin{equation}}
\def\eeq{\end{equation}}
\def\bea{\begin{eqnarray}}
\def\eea{\end{eqnarray}}
\def\bd{\begin{displaymath}}
\def\ed{\end{displaymath}}

\newcommand{\Cint}{\int\kern-10.5pt-\kern7pt}

\newcommand{\be}{\begin{equation}}
\newcommand{\ee}{\end{equation}}
\newcommand{\bal}{\begin{align}}
\newcommand{\eal}{\end{align}}

\newcommand\fverbit{\egroup\item[\fbox{\unhbox\pippobox}]}
\newbox\pippobox

{

\def\be{\begin{equation}}
\def\ee{\end{equation}}
\def\bea{\begin{eqnarray}}
\def\eea{\end{eqnarray}}

\begin{document}

\begin{titlepage}
\setcounter{page}{1}

\vskip 5cm

\begin{center}

\vspace*{3cm}

{\Large ON THE NONEQUILIBRIUM DYNAMICS \\[8pt] OF GRAVITATIONAL ALGEBRAS}

\vspace{15mm}

{\large Michele Cirafici}
\\[6mm]
\noindent{\em Dipartimento di Matematica, Informatica e Geoscienze, \\ Universit\`a di Trieste, Via A. Valerio 12/1, I-34127,
 \\ Institute for Geometry and Physics \& INFN, Sezione di Trieste,  Trieste, Italy 
}\\ Email: \ {\tt michelecirafici@gmail.com}

\vspace{15mm}

\begin{abstract}
\noindent 
We explore nonequilibrium features of certain operator algebras which appear in quantum gravity. The algebra of observables in a black hole background is a Type $\rm{II}_\infty$ von Neumann algebra. We discuss how this algebra can be coupled to the algebra of observable of an infinite reservoir within the canonical ensemble, aiming to induce nonequilibrium dynamics. The resulting dynamics can lead the system towards a nonequilibrium steady state which can be characterized through modular theory. Within this framework we address the definition of entropy production and its relationship to relative entropy, alongside exploring other applications.
\end{abstract}

\vspace{15mm}

\today

\end{center}
\end{titlepage}


\tableofcontents


\section{Introduction}

The Quantum Extremal Surface prescription \cite{Engelhardt:2014gca} has played an important role in recent advancements toward deriving the Page curve for evaporating black holes \cite{Almheiri:2019psf,Penington:2019npb,Penington:2019kki,Almheiri:2019qdq}. More broadly, this underscores the significance of a comprehensive understanding of entropy within the context of quantum gravity. A crucial step in unraveling the mechanisms of the AdS/CFT correspondence lies in precisely determining how information about bulk degrees of freedom is encoded on the boundary.

Recently the use of operator algebras has emerged as a promising tool for elucidating the connection between quantum gravity, entropy and information \cite{Papadodimas:2012aq,Papadodimas:2013jku,Leutheusser:2021qhd,Leutheusser:2021frk}. In particular a proper consideration of gravitational dynamics in a black hole background naturally leads to a Type $\rm{II}$ von Neumann algebra \cite{Witten:2021unn,Chandrasekaran:2022eqq}. These results have been extended to various directions, such as other spacetimes \cite{Chandrasekaran:2022cip,Bahiru:2022mwh,Gomez:2023wrq,Kudler-Flam:2023qfl} or subregions thereof \cite{Leutheusser:2022bgi,Jensen:2023yxy,AliAhmad:2023etg}, diverse setups \cite{Penington:2023dql,Gesteau:2023hbq,Kolchmeyer:2023gwa,Aguilar-Gutierrez:2023odp,Bahiru:2023ify,Krishnan:2023fnt,Engelhardt:2023xer} and the realm of quantum chaos \cite{Furuya:2023fei,Gesteau:2023rrx,Ouseph:2023juq}. Recent articles reviewing aspects of the theory relevant to this article include \cite{Witten:2018zxz,Witten:2021jzq,Sorce:2023fdx,Sorce:2023gio}. The resulting gravitational algebras appear to encode most of the relevant properties anticipated in quantum gravity.

Several processes involving gravity, such as black hole evaporation, occur outside of equilibrium. While equilibrium thermodynamics has been instrumental in understanding black hole physics and gravity in general, certain processes necessitate departure from this regime. The crucial role played by von Neumann algebras in recent developments provides an avenue to connect with the formalism of nonequilibrium statistical mechanics. 

In this paper we take a first modest step in this direction by adapting the general setup used to study nonequilibrium quantum statistical mechanics, as reviewed in \cite{RuelleRev,JP1,JP0}, to the gravitational algebras which appear in the context of holography. We achive this first by coupling the gravitational algebras to an external reservoir.  The implementation of this coupling requires the gravitational algebras to be associated with the canonical ensemble formalism of \cite{Witten:2021unn}. Such gravitational algebras are Type $\rm{II}_\infty$ algebras which arise from the crossed product of Type $\rm{III}_1$ algebras. Physically this crossed product corresponds to incorporating $1 / N$ corrections in the boundary theory.  While coupling the boundary theory to a reservoir involves a straightforward modification of the relevant Hamiltonians, including $1 / N$ corrections is more subtle, necessitating the study of KMS structures and their deformations, as well as aspects of Tomita-Takesaki theory.

Once the coupling is established, the system can transition to a new equilibrium state or, more intriguingly, approach a nonequilibrium steady state (NESS). The characterization of NESS is challenging, but we implicitly describe it in terms of the system's dynamics. In particular we can relate explicitly the entropy production during the evolution to the relative entropy between the relevant state and a reference state. Finally we discuss how we expect our results to be useful in the study of evaporating black holes and of quantum chaos.

The purpose of this note is to establish an algebraic formalism, which will be used elsewhere. We aim for generality, somewhat overlooking practical applications. The latter heavily depend on specific details chosen for the interaction term between the boundary theory and the reservoirs. 

The main achievements of this note revolve around extending various results from nonequilibrium statistical mechanics to gravitational algebras. In the literature, these results are tailored for finite-dimensional systems, typically characterized by a type $I_n$ algebra, interacting with an infinite reservoir. While we will see that extending these findings to type $II_\infty$ algebras can be straightforward in some cases, it often requires a more intricate approach. For example this will be the case when computing the modular operator of the system coupled to the reservoir or delving into the analysis of nonequilibrium steady states and entropy production for classical-quantum states.

This note is organized as follows. Sections \ref{ModStructures} and \ref{GravAlg} are reviews of certain aspects of operator algebras and the construction of gravitational algebras that are relevant to this paper. In Section \ref{NEdyn} we explore the nonequilibrium dynamics of these gravitational algebras arising from coupling to reservoirs, and in Section \ref{ep}, we discuss entropy production. In both Sections we briefly discuss the analog results for finite dimensional systems before generalizing them to the case at hand. Finally in Section \ref{applications} we briefly touch upon some applications, before ending with our conclusions.

\section{Modular structures} \label{ModStructures}

In this Section we briefly review certain aspects of quantum dynamical systems and modular theory towards the study of systems out of equilibrium.

\paragraph{Quantum dynamical systems.} The thermodynamic limit of quantum systems can be conveniently described using operator algebras. In this formalism a ($W^*$) quantum dynamical system is a pair $(\cA , \alpha)$ where $\cA$ is a von Neumann algebra and $\real \ni t \rightarrow \alpha^t$ a one parameter group of $*$-automorphisms of $\cA$. The dynamics $\alpha$ is formally defined for $\sfa \in \cA$ as 
\be
\alpha^t \left( \sfa \right) = \sum_{m=0}^\infty \frac{t^m}{m!} \delta^m \sfa = \e^{t \delta} \sfa
\ee
where $\delta$ is the infinitesimal generator of $\alpha$. The generator is a derivation, with $\delta (\sfa \, \sfb) = \delta (\sfa) \, \sfb + \sfa \, \delta (\sfb)$ and $\delta (\sfa^\dagger) = \delta (\sfa)^\dagger$.

For example if $\cH$ is a finite dimensional Hilbert space and $\cB (\cH)$ denotes the algebra of bounded operators on $\cH$, then given $H$ a self-adjoint operator,
\be
\alpha^t (\sfa) = \e^{\ii t H} \, \sfa \, \e^{- \ii t H}
\ee
is a dynamics generated by $\delta (\sfa) = \ii \left[ H , \sfa \right]$.

A state $\omega$ on the algebra $\cA$ is $\alpha$-invariant if $\omega \circ \alpha^t = \omega$ for all $t \in \real$ (by which we mean $\omega (\alpha^t (\sfa)) = \omega (\sfa)$ for every operator $\sfa \in \cA$).  To an invariant state $\omega$ on the quantum dynamical system $\left( \cA , \alpha \right)$ the GNS construction associates the triple 
$\left( \cH_\omega , \pi_\omega , \Omega_\omega \right)$ and the dynamics is generated by a one-parameter group $t \longrightarrow U_{\omega} (t)$ of unitary operators
\be
\pi_\omega \left( \alpha^t (\sfa) \right) =U_\omega (t) \, \pi_\omega (\sfa) \, U_\omega (t)^\dagger \, 
\ee
which is unique for a given $\alpha$.

We can define a perturbed dynamics starting from a perturbation $V \in \cA$ via the generator
\be
\delta_V = \delta + \ii \left[ V , \ \cdot \ \right]
\ee
and setting $\alpha^t_V = \e^{t \delta_V}$. The perturbed dynamics is described by the Dyson expansion
\be
\alpha^t_V (\sfa) = \alpha^t (\sfa) + \sum_{n=1}^\infty \int_0^{t} \dd t_1 \int_0^{t_1} \dd t_2 \cdots \int_0^{t_{n-1}} \dd t_{n} \ii \left[ \alpha^{t_n} (V) , \ii \left[ \cdots , \ii \left[ \alpha^{t_1} (V) , \alpha^t (\sfa) \right] \cdots \right] \right] \, .
\ee

\paragraph{Quantum KMS states.} A state $\omega$ on $\left( \cA , \alpha \right)$ is $\left( \alpha , \beta \right)$-KMS for some inverse temperature $\beta \in \real$ if $\forall  \, \sfa , \sfb \in \cA$ there exists a function $\cF_{\beta} (\sfa , \sfb ; z)$ which is analytic in the strip
\be
S_\beta = \left\{ z \in \complex \, \vert \, 0 < \mathrm{Im} (z \, \sign \, \beta) < |\beta| \right\},
\ee
continuous on its closure and such that the following two conditions hold
\begin{itemize}
\item $\cF_\beta (\sfa , \sfb ; t) = \omega (\sfa \, \alpha^t (\sfb))$,
\item $\cF_\beta (\sfa , \sfb ; t + \ii \beta) = \omega (\alpha^t (\sfb) \, \sfa)$ \, ,
\end{itemize}
on its boundary.
Note that if $\omega$ is $(\alpha^t , \beta)$-KMS, then it is also $(\alpha^{k t} , \beta / k)$-KMS. On the other hand KMS states at different temperatures but with the same dynamics are not simply related. 


The KMS conditions are equivalent to 
\be
\omega ( \sfb \, \sfa) = \omega (\sfa \, \alpha^{\ii \beta} (\sfb))
\ee
which can be taken as characterization of KMS states. In particular this condition implies that a $(\alpha , \beta)$-KMS state is $\alpha$-invariant. 

In the example of a finite dimensional Hilbert space $(\cB (\cH) , \alpha)$, with $\alpha$ given in terms of a Hamiltonian $H$, there is a unique thermal equilibrium state $\omega$ given by
\be
\omega (\sfa) = \frac{1}{Z} \Tr \left( \e^{-\beta H} \sfa \right)
\ee
with $Z = \Tr \left( \e^{-\beta H}\right)$. Then
\be \label{KMSft}
\omega (\sfa \alpha^{\ii \beta} (\sfb)) = \frac{1}{Z} \Tr \left( \e^{-\beta H} \sfa \e^{\ii (\ii \beta) H} \sfb  \e^{- \ii (\ii \beta) H} \right) 
= \frac{1}{Z} \Tr \left( \e^{-\beta H} \sfb \sfa \right) = \omega (\sfb \sfa) \, .
\ee

A crucial result in Araki's theory of perturbation of KMS structures is that there is a one to one correspondence between the set of $(\alpha , \beta)$-KMS states and the set of $(\alpha_V , \beta)$-KMS states.

In the course of the paper we will sometime encounter states which corresponds to vectors in some Hilbert space. We will denote the functional and the vector with the same letter, but lower and upper case respectively; for example the functional $\varphi$ and its vector $\Phi$. This notation will descend to the crossed product where states are denoted with hats, for example the functional $\widehat{\varphi}$ and the vector $\widehat{\Phi}$. 

\paragraph{Tomita-Takesaki theory.} Consider a von Neumann algebra $\cA$ acting on a vector space $\cH$. Recall that a vector $\Psi \in \cH$ is cyclic if $\cA \Psi$ is dense in $\cH$ and separating if $\sfa \Psi = 0$ for some $\sfa \in \cA$ implies $\sfa = 0$. A vector which is both cyclic and separating is called modular. The relation between modular theory and finite temperature physics is that given a dynamics $(\cA , \alpha)$ any $(\alpha , \beta)$-KMS state is modular.

If $\Psi$ is a modular vector for $\cA$, then the map $\sfa \Psi \longrightarrow \sfa^\dagger \Psi$ determines an anti-linear involution $S$. Its polar decomposition $S = J \, \Delta^{1/2}_\Psi$ define the modular conjugation $J$ and the modular operator $\Delta_\Psi$. The cornerstone of Tomita-Takesaki theory is the result:
\begin{itemize}
\item $J \cA J = \cA'$, where $\cA'$ is the commutant of $\cA$,
\item $\Delta_\Psi^{- \ii t} \cA \Delta_\Psi^{\ii t} = \cA$ for any $t \in \real$
\end{itemize}
In particular $\sigma^t (\sfa) = \Delta_\Psi^{-\ii t} \sfa \Delta_\Psi^{\ii t}$ defines a $*$-automorphism of $\cA$ which is called the modular group. 

The modular group is deeply related to the KMS condition. An equivalent characterization of the modular operator is 
\be
\braket{\Psi \vert \sfa \sfb \vert \Psi} = \braket{\Psi \vert \sfb \, \Delta_\Psi \, \sfa \vert \Psi} \, .
\ee
Consider for example a finite dimensional case. The modular operator defines the modular hamiltonian $H$ as $\Delta_\Psi = \e^{- \beta H}$. Suppose the system is in thermal equilibrium so that the modular operator is associated to a Gibbs distribution. Then if $\sfa (t ) = \alpha^t (\sfa) =  \e^{\ii t H} \sfa \e^{- \ii t H}$ denotes the time evolved operator in the Heisenberg picture, we see that
\begin{align}
\braket{\Psi \vert \sfa (t) \, \sfb \vert \Psi} &= \braket{\Psi \vert \sfb \, \Delta_\Psi \, \sfa(t) \vert \Psi} 
=  \braket{\Psi \vert \sfb \, \e^{\ii (t + \ii \beta) H} \, \sfa \e^{- \ii (t + \ii \beta) H } \vert \Psi} = \braket{\Psi \vert \sfb \, \sfa(t + \ii \beta) \vert \Psi} \, ,
\end{align}
since $H \ket{\Psi} = 0$. This shows that the state $\Psi$ out of which the modular group is constructed is $(\alpha^t , \beta)$-KMS\footnote{This simple example also helps to set up our conventions. In the following we will deal with modular hamiltonians $H$ defined via $\Delta = \e^{- H}$. Then the associated state is KMS with inverse temperature $\beta = +1$. Most of the mathematics literature use the opposite convention, defining $\Delta = \e^{H}$, so that the associated state is KMS with inverse temperature $\beta = -1$. }. 

The above statement also holds in the infinite dimensional setting, where it is a very non-trivial condition. It implies certain analytical properties of the correlators. In essence it says that every state on a von Neumann algebra determines an automorphism (a ``time'' flow) on the algebra, for which the original state is KMS. Furthermore thanks to the Connes' cocycle  theorem, two automorphisms induced by two different states via the associated modular flow are equivalent, by which we mean they are related by the composition of another automorphism. We refer the reader to \cite{Sorce:2023gio} for a nice discussion of these matters and accessible proofs of the relevant statements.

Only in special conditions such flows have a direct physical interpretation. In the example above the KMS condition characterizes thermal equilibrium. In thermal equilibrium the modular automorphism group represents time evolution. This establishes the physical link between the modular hamiltonian and the automorphism group which determines the dynamics. This is the familiar statement in equilibrium statistical mechanics that the statistical weight of a thermal state is determined by the evolution operator. See for example \cite{RuelleRev} for a discussion.


\paragraph{Relative entropy.} Similarly given two states $\Psi$ and $\Phi$ one can introduce the relative Tomita operator $S_{\Phi \vert \Psi}$ whose action is
\be
S_{\Phi \vert \Psi} \, \sfa \, \ket{\Psi} = \sfa^\dagger \ket{\Phi} \, .
\ee
The relative modular operator is then defined as $\Delta_{\Phi \vert \Psi} = S^\dagger_{\Phi \vert \Psi} \, S_{\Phi \vert \Psi}$, and the relative modular hamiltonian is given by $h_{\Phi \vert \Psi} = - \log \Delta_{\Phi \vert \Psi}$. The relative modular operator can be used to define the relative entropy
\be
S \left( \Psi \Vert \Phi \right) = - \braket{\Psi \vert \log   \Delta_{\Phi \vert \Psi}  \vert \Psi} \, .
\ee
For example in a finite dimensional setting we can represent the state $\Psi$ with a density matrix $\rho$ and the state $\Phi$ with a density matrix $\sigma$. In this case the 
the relative entropy assumes the familiar form
\be
S \left( \Psi \Vert \Phi \right) = \Tr \, \rho \left( \log \rho - \log \sigma \right) \, .
\ee

\section{Gravitational algebras from holography} \label{GravAlg}

The work of  \cite{Leutheusser:2021qhd,Leutheusser:2021frk} shows the natural emergence of a certain algebra in the large $N$ limit of thermal correlators in $\cN=4$ SYM. In such limit the only non vanishing thermal correlator of subtracted single-trace operators of the form $\cW = \Tr W - \langle \Tr W \rangle_\beta$ is the two point function. Operators with this behaviour define  generalized free fields. The set of non-central operators of this form in the large $N$ limit form a certain von Neumann algebra. This algebra acts naturally on a separable Hilbert space defined via the GNS construction, starting from a vector $\Psi$. The inner product on this space is determined by the large $N$ limit of thermal correlators at a certain temperature $T = 1 / \beta$. When such temperature is above the Hawking-Page temperature the algebra is a von Neumann algebra $\cA_{0,R}$ of Type $\mathrm{III}_1$. Above such temperature the state $\Psi$ is naturally identified with the thermofield double state
\be
\Psi_{TFD} = \frac{1}{\sqrt{Z}} \ \sum_i \exp (- \beta E_i / 2) \ket{\overline{E}}_L \otimes \ket{E}_R
\ee
which physically represents a purification of the thermal density matrix
\be
\rho_\beta = \Tr  \ket{\Psi_{TFD}} \bra{\Psi_{TFD}} = \frac{1}{Z (\beta)} \sum_i \e^{- \beta E_i} \ket{E}_r \bra{E}_r =\frac{1}{Z (\beta)} \e^{- \beta H}
\ee
whenever this can be defined. In particular it is a pure state and its correlation functions are thermal. The Hilbert space constructed in this way has a well defined large $N$ limit.

In the AdS/CFT correspondence the dual of the thermofield double state above the Hawking-Page temperature is the two-sided eternal black hole \cite{Maldacena:2001kr}. The algebra $\cA_{R,0}$ is associated with one of the boundaries and its commutant $\cA_{L,0} = \cA_{R,0}'$ is a von Neumann algebra of Type $\mathrm{III}_1$ associated with the other boundary. The algebras are respectively dual to the bulk algebras $\cA_{r,0}$ and $\cA_{l,0}$ which describe bulk quantum fields in the left and right exteriors of the eternal black hole \cite{Leutheusser:2021qhd,Leutheusser:2021frk}. 

The two algebras $\cA_{R,0}$ and $\cA_{L,0}$ are factors, which means that their center consists only of $c$-numbers. This is a consequence of the fact that the construction of the algebras involves only noncentral single-trace operators. Central single-trace operators form a finite dimensional set and are associated with conserved charges. Among those are the boundary Hamiltonians $H_R$ and $H_L$ which are related to the black hole mass via the duality. However the Hamiltonians do not have a large $N$ limit but only their difference $\widehat{H} = H_R - H_L$ does. Indeed $\widehat{H}$ annihilates the thermofield double state. This operator is dual to $\widehat{h}$, the conserved charge associated with the killing vector field that generates time translations in the eternal black hole background, via $\widehat{h} = \beta H$.

To obtain a sensible large $N$ limit one defines the rescaled and subtracted operators
\be
U_L = \frac{H_L - \braket{H_L}_\beta}{N} \, , \qquad U_R = \frac{H_R - \braket{H_R}_\beta}{N} \, .
\ee 
These operators have a large $N$ limit and they are central when $N=\infty$. Moreover $U_R - U_L = \widehat{H} / N$ and therefore coincide in the strict $N=\infty$ limit. In this case we can drop the labels and simply call these operators $U$; $U$ is a central operator for both $\cA_{R,0}$ and $\cA_{L,0}$.

Therefore one can incorporate the central generator $U$ by simply tensoring the original algebras with the algebra of bounded functions of $U$. The algebra $\cA _R = \cA_{R, 0 } \otimes \cA_U$ acts now on $\widehat{\cH}_{TFD} = \cH_{TFD} \otimes L^2 (\real)$ and its elements look like
\be
\widehat{\sfa} = \int_{- \infty}^{\infty} \dd u \ \sfa (u) \e^{\ii u \, U}
\ee
for $\sfa \in \cA_{R,0}$. Now the thermofield double can be written as $\widehat{\Psi}_{TFD} = \Psi_{TFD} \otimes g^{1/2} (U)$, with $g$ a Gaussian function. State of this kind are called classical-quantum states. A similar story holds for $\cA _L = \cA_{L, 0 } \otimes \cA_U$. In the large $N$ limit $\cA_L$ and $\cA_R$ are still Type $\rm{III}_1$ algebras, but they are not factors due to the presence of a non-trivial center, generated by $U$. 

It was shown in \cite{Witten:2021unn} that taking into account $1 / N$ corrections substantially improves the picture and turn the algebras into von Neumann algebras of Type $\rm{II}_\infty$. These kind of algebras are quite different from the Type $\rm{III}$ algebras which appear in quantum field theory. They do not have irreducible representations, which means that they do not describe microstates, but have the advantage that one can define traces and density matrices, and therefore von Neumann entropies. 

The relevant construction is the so-called crossed product. We will now review in some detail its definition and application to the case at hand following \cite{Witten:2021unn,Chandrasekaran:2022eqq}, since we will use this construction repeatedly in Section \ref{NEdyn}. Abstractly the crossed product is defined as follows.  Let $\cA$ be an algebra acting on a Hilbert space $\cH$. The self-adjoint operator $T$ generates a one parameter group of automorphisms by $\e^{\ii s \, T} \cA \e^{-\ii s \, T}$. The crossed product $\cA \rtimes \real$ is defined as the algebra acting on the Hilbert space $\widehat{\cH} = \cH \otimes L^2 (\real)$ generated by $\sfa \otimes 1$ and $\e^{\ii s \, T} \otimes \e^{\ii s \, X}$. Here $L^2 (\real)$ is the space of square summable functions of the auxiliary variable $X$.

If $T$ is an inner automorphism the crossed product reduces to the ordinary tensor product. The interesting case is when the automorphism generated by $T$ is outer. In this case if the algebra $\cA$ is a $\rm{III}_1$ factor, then the crossed product algebra $\cA \rtimes \real$ is a $\rm{II}_{\infty}$ factor and the automorphism generated by $T$ becomes inner. This is precisely our case. The thermofield double state $\Psi_{TFD}$ is associated to the modular operator $\Delta = \e^{- \beta \widehat{H}}$ with $\widehat{H} \ket{\Psi_{TFD}}  = 0$. Tomita-Takesaki theory implies that the automorphism generated by $\beta \widehat{H}$ is outer for Type $\rm{III}$ algebras, as is our case. 

When we consider also $1 / N$ corrections, the two operators $U_L$ and $U_R$ become distinct, since $U_R = U_L + \beta \widehat{H} / N$. The operator $U_L$ still commutes with $\cA_{R,0}$. We can therefore take $X = \beta N U_L$. The crossed product algebra $\cA_R$ is then the algebra that acts on $\widehat{\cH} = \cH \otimes L^2 (\real)$ obtained by adjoining to $\cA_{R,0}$ bounded functions of $\beta \widehat{H} + X$. This combination is morally $H_R$, which is not a well defined operator in the large $N$; the crossed product construction provides a mean to include this operator in the algebra in a meaningful way.

The fact that  $\cA_R$ is obtained via the crossed product by the modular automorphism group seems to indicate that the full construction depends on the vector $\Psi_{TFD}$. However this is not the case, as was proven in \cite{Witten:2021unn} by means of the Connes' cocycle. Therefore henceforth we will drop the subscript $TFD$ and denote by $\Psi$ a generic cyclic and separating vector and by $\Delta_\Psi$ its modular operator.

For classical-quantum states of the form $\widehat{\Psi} = \Psi \otimes g (X)^{1/2}$ we can write down explicitly the modular operator \cite{Witten:2021unn}
\be
\widehat{\Delta}_{\widehat{\Psi}} = \Delta_{\Psi} \, g (\beta \widehat{H} + X) \, g (X)^{-1} = K \widetilde{K} \, ,
\ee
where $K \in \cA \rtimes \real_{\Psi}$ and $\widetilde{K} \in \left( \cA \rtimes \real_{\Psi} \right)'$ are given by
\begin{align}
K &= \e^{- \left(\beta \widehat{H} + X \right)} g (\beta \widehat{H} + X) \, , \\ 
\widetilde{K} &= \e^X g(X)^{-1} \, .
\end{align}
In other words the modular operator factorizes. In the Type $\rm{I}$ algebras, which appear in ordinary quantum mechanics, the Hilbert space of a bipartite system factorizes; similarly the modular operator can be expressed as product of ordinary density matrices. For Type $\rm III$ algebras,  which appear in quantum field theory, neither the Hilbert space nor the modular operator factorize. The Type $\rm II$ case is an intermediate situation where the Hilbert space does not factorize but the modular operator does.

This factorization is important because we can use it to define the trace of an operator $\widehat{\sfa} \in \cA_R$ as
\be \label{trK-1}
\tr \, \widehat{\sfa} = \braket{\Psi | \widehat{\sfa} K^{-1} | \Psi}
\ee
which is cyclic \cite{Witten:2021unn}. Note that this is defined up to a factor of $\e^{-c}$ which comes from the rescaling of $K$. For example if 
\be
\widehat{\sfa} = \int_{- \infty}^{\infty} \dd u \ \sfa (u) \e^{\ii u \, U}
\ee
then 
\be \label{traceII}
\tr \, \widehat{\sfa} = \int_{- \infty}^{\infty} \dd X \, \e^X \, \braket{\Psi | \sfa (X) | \Psi} \, .
\ee
This algebra is a factor of Type $\rm{II}_\infty$ and therefore the trace is not defined on all the elements, but only on those such that the integral in \eqref{traceII} is convergent. 

This definition of a trace is however not completely satisfactory. The appearance of the operator $X$ in the exponent means that there is a factor of $N$ in the exponent. As a result these traces cannot be expressed as formal power or Laurent series in $1 / N$ but must be considered as formal functions of $N$. Luckily this problem can be circumvented in the computations of the entropies \cite{Chandrasekaran:2022eqq}. On the other hand using \eqref{traceII} as a formal expression can in many cases help the intuition; therefore we will still employ it in the following.

If we accept that the above formula define a suitable trace, we can also introduce density matrices. Recall that an element $\rho_{\widehat{\Phi}} \in \cA_{R}$ is a density matrix for a state $\widehat{\Phi}$ if for all operators $\widehat{a} \in \cA_{R}$ we have 
\be 
\tr \, \widehat{\sfa} \, \rho_{\widehat{\Phi}} = \braket{\widehat{\Phi} | \widehat{\sfa} | \widehat{\Phi}} \, . 
\ee
This density matrix is normalized to $1$ but scales as $\rho_{\widehat{\Phi}} \longrightarrow \e^{-c} \rho_{\widehat{\Phi}}$ with the trace. Finally existence of density matrices and of traces allow us to define the von Neumann entropy $S (\rho) = - \Tr \, \rho \log \rho$. For a state $\widehat{\Phi}$ described by a density matrix $\rho_{\widehat{\Phi}}$ the von Neumann entropy is defined as
\be
S (\widehat{\Phi})_{\cA_R} = - \braket{\widehat{\Phi} \vert \log \rho_{\widehat{\Phi}} \vert \widehat{\Phi} } \, .
\ee
It follows from the above discussion that such an entropy is defined only up to an additive constant and therefore only entropy differences are meaningful. Similarly one can also define Renyi entropies. 

For example taking the classical-quantum state $\widehat{\Psi} = \Psi \otimes g(X)^{1/2}$, it follows from \eqref{trK-1} that $\rho_{\widehat{\Psi}} = K$ and therefore
\be \label{vnPsi}
S (\widehat{\Psi})_{\cA_R} = \int_{- \infty}^{\infty} \dd X \left( X \, g(X) - g(X) \, \log X \right) \, .
\ee

As we have discussed, traces and density matrices are not suitably defined for a Laurent expansion in powers of $1 / N$. Nevertheless entropies are \cite{Chandrasekaran:2022eqq}. One can define entropies in the canonical ensemble bypassing the construction of traces and density matrices as the expectation values of an operator which generates the modular flow. Indeed such an operator, morally the logarithm of a density matrix, generates an inner automorphism in Type $\rm{II}$ algebras. 

Consider a classical-quantum state $\widehat{\Phi} = \Phi \otimes f (X)^{1/2}$. One can introduce the operator  \cite{Chandrasekaran:2022eqq}
\be
h_{\widehat{\Phi}} = - \frac{1}{N} \log \rho_{\widehat{\Phi}}
\ee
even if the expression $\rho_{\widehat{\Phi}}$ is only formal. One finds that
\be
N h_{\widehat{\Phi}} = X + h_{\Phi \vert \Psi} - \log | f (U_R) | + \alpha (U_R) + N S_0 \, ,
\ee
which gives the entropy
\be \label{EntropyPhi}
S (\widehat{\Phi})_{\cA_R} = N \beta \braket{U_R} + N S_0 - S  \left( \Phi \Vert \Psi \right) - \braket{\log | f (U_R) |} + \braket{\alpha (U_R)} \, .
\ee
Here we have used the relation $h_\Psi - h_{\Phi | \Psi} = h_{\Psi | \Phi} - h_{\Phi}$ which follows from the properties of the Connes' cocycle. The function $\alpha$ can be fixed by going to the next order in the $1 / N$ expansion as
\be
\alpha (U_R) = - \frac{N^2}{T^2 \, C_{\rm{BH}}} \frac{U^2_R}{2} + \rm{const} \, ,
\ee
where $C_{\rm{BH}}$ is the black hole heat capacity. In \eqref{EntropyPhi} both the constant $ N S_0 $ and the function $\braket{\alpha (U_R)}$ are state independent and drop out when computing entropy differences.

Alternatively one can pass to the microcanonical ensemble, where traces and density matrices can be defined without formal arguments \cite{Chandrasekaran:2022eqq}. However for the purpose of studying the nonequilibrium dynamics we will continue to use the canonical ensemble. 

\section{Nonequilibrium dynamics} \label{NEdyn}

In this section we discuss the nonequilibrium dynamics of the gravitational algebras introduced in Section \ref{GravAlg}. We start with a brief discussion on the operator formalism in nonequilibrium statistical mechanics. Afterwards we couple the gravitational algebras to external reservoirs. We discuss separately the case where the perturbed system approaches a KMS equilibrium state or a genuinely nonequilibrium steady state. In the KMS case we use the relation between statistical weights in thermal states and the quantum dynamics to determine an automorphism of the algebra which plays the role of time evolution. When discussing nonequilibrium physics we will follow \cite{RuelleRev,JP1} closely and when needed generalize their results to algebras obtained from the crossed product.

\subsection{Generalities}

At this stage we have an operator algebra which describes quantum gravity effects in a black hole background. The black hole is in thermal equilibrium with its surroundings. We want to introduce a perturbation to drive the system out of equilibrium. In principle this can be done in several ways. In classical nonequilibrium statistical mechanics there are two principal methods.

In the first scenario one starts with an Hamiltonian system with a large, but still finite, number of degrees of freedom. The system is taken out of equilibrium, for example by time dependent interactions, but still constrained to live within a fixed energy interval by a thermostat. This scenario corresponds to a statistical description in terms of the microcanonical ensemble.

In the second scenario we couple the original system to a number of reservoirs in thermal equilibrium at different temperatures and we allow for the exchange of energy between the original system and the reservoirs. This framework corresponds to the canonical ensemble. 

It is this second scenario which we choose to work within. The reason is that it is easier from the conceptual point of view. We should however stress that our system is infinite dimensional to begin with. Furthermore the canonical gravitational algebras, as discussed in Section \ref{GravAlg}, are less understood than their microcanonical counterpart. Ultimately one expects that all the  ensembles become equivalent as the number of degrees of freedom goes to infinity, but there are certain aspects which still have to be clarified. For the present being we will ignore subtleties in the definition of the canonical ensemble, like the difficulty in defining traces in the $1/N$ expansion, leaving a more in depth discussion for the future.

\subsection{Close to and far from equilibrium}

Equilibrium thermodynamics is characterized by the observation that macroscopic equilibrium states are operationally definable in terms of macroscopic quantities, such as temperature and entropy, that are non mechanical. In equilibrium there is no explicit time dependence and no more reference to the dynamics after it has been used to define the microscopic ensembles. 

Outside of the equilibrium regime, temperature and entropy may vary, with time and eventually with the position. The simples class of nonequilibrium states are the nonequilibrium steady states (NESS). From a phenomenological point of view nonequilibrium steady states arise for example when bringing in contact two isolated systems in equilibrium at different temperature. Before the whole system thermalizes, there will be a regime characterized, for example, by a steady flow of heat from one system to another. The precise details depend on the interaction between the two systems. Nonequilibrium steady states are simply defined as limits of initial states under time evolution, if such limits exist.

Consider a quantum dynamical system $(\cA , \alpha)$ which is in a $\alpha$-invariant state $\omega$. We imagine perturbing the system with a self-adjoint operator $V = V^{\dagger} \in \cA$ and denote by $\alpha_V$ the perturbed dynamics. Let $\left( \cH_\omega , \pi_\omega , \Omega_\omega \right)$ be the GNS representation of the algebra $\cA$ associated with the state $\omega$. When no confusion is expected to arise, we will drop the label $\omega$ from the relevant quantities and omit the map $\pi_\omega$. A state $\eta$ is called $\omega$-normal if for every operator $\sfa \in \cA$ it can be represented by using a density matrix, that is $\eta (\sfa) = \Tr \rho \, \pi_\omega (\sfa)$.

Nonequilibrium steady states are then defined as limits of states under the perturbed evolution
\be \label{NESSdef}
\omega_+ (\sfa) = \lim_{t_k } \frac{1}{t_k} \int_{0}^{t_k} \, \omega \circ \alpha_V^t (\sfa) \, \dd t =  \lim_{t_k } \frac{1}{t_k} \int_{0}^{t_k} \braket{\Omega_\omega \vert \alpha_V^t (\sfa) \vert \Omega_\omega} \, \dd t \, ,
\ee
if such limit exists for a divergent sequence $\{ t_k \}_{n \in \zed_+}$ and for all $\sfa \in \cA$. Just like in the case of thermal equilibrium, where a system can have several KMS states, nonequilibrium steady states need not be unique. We denote by $\Sigma^+_V (\omega)$ the set of nonequilibrium steady states which can be reached from $\omega$ due to the perturbation $V$. Note that in particular $\omega_+$ could be a KMS state. Indeed Araki's theory of perturbations of KMS states gives a general framework to understand when this is the case, see for example \cite{BR2}. Similarly one can define $\Sigma^-_V (\omega)$, in the far past. Note that a NESS is necessarily an $\alpha_V$-invariant state.

We also expect that the limit is independent of the initial state in the sense that initial states which are not too far apart, under weak ergodic hypotheses, will converge to the same limit. For example it is enough that \cite{JP1}
\be
\lim_{T \longrightarrow \pm \infty} \frac{1}{T} \int_0^T \eta \left( \left[ \alpha^t_V (A) , B \right] \right) \ \dd t = 0
\ee
for all $\eta \in \cN_\omega$. This condition of \textit{asymptotic abelianness} is an ergodic condition and in general very difficult to prove rigorously. It implies that in the long-time limit observables become effectively non-interacting. For example it would be satisfied if the system thermalizes. If this is the case then all $\eta \in \cN_\omega$ tend to the same limit and $\Sigma^{\pm}_V (\eta) = \Sigma^\pm_V (\omega)$. We expect this condition to be satisfied for generalized free fields.

Any state can be decomposed into a normal part and a singular part, $\eta = \eta_n + \eta_s$ with $\eta_n \in \cN_\omega$ and $\eta_s$ is defined as the remaining part. The rationale behind this decomposition is that the entropy production of a purely normal and $\alpha_V$-invariant state is always zero, as we will see more in detail in Section \ref{ep}. 

More in detail let us suppose that we start from a KMS state $\omega$ in a quantum dynamical system $(\cA , \alpha)$. After a weak perturbation the system will eventually set in a NESS. Under rather general conditions if the perturbation is sufficiently weak we expect that this NESS has the form of $\omega_V$, a KMS state for the perturbed dynamics $(\cA , \alpha_V)$. In this case, as we will see momentarily, Araki's theory gives a precise expression for $\omega_V$.

\paragraph{Example.} Consider again a finite dimensional isolated system governed by an hamiltonian $H_S$ acting on a Hilbert space $\cH$. Introduce now a perturbation $H_S + V$, so that the perturbed dynamics is given by
\be
\alpha_V (\sfa) = \e^{\ii t (H_S + V )} \sfa \e^{- \ii t ( H_S + V)}
\ee
The system is originally in the $(\alpha , \beta)$-KMS Gibbs state
\be
\omega (\sfa) = \frac{\Tr_\cH \e^{- \beta H_S} \sfa}{\Tr_\cH \e^{- \beta H_S}} \, ,
\ee
and the state  
\be
\omega_V (\sfa) =  \frac{\Tr_\cH \e^{- \beta (H_S + V)} \sfa}{\Tr_\cH \e^{- \beta (H_S + V)}}
\ee
is the unique $(\alpha_V , \beta)$-KMS state associated with the perturbed dynamics. Note however that by introducing
\be
\Gamma^V_{t} = \e^{\ii t (H_S + V)} \, \e^{- \ii t H_S}
\ee
and analytically continuing it to $\ii \beta/2$, we can write
\be
\omega_V (\sfa) = \frac{\omega \left( (\Gamma^V_{\ii \beta/2})^\dagger \sfa (\Gamma^V_{\ii \beta/2}) \right)}{\omega \left(  (\Gamma^V_{\ii \beta/2})^\dagger (\Gamma^V_{\ii \beta/2}) \right)} \, ,
\ee
where the adjoint is needed to have the correct ordering of the operators. In particular it follows that 
\be
\Omega_{\omega_V} = \frac{\e^{\beta/2 \, (H_S + V)} \Omega_\omega}{\Vert \e^{\beta/2 \, (H_S + V)} \Omega_\omega \Vert} \, .
\ee
Since $\Gamma^V_t$ can be expanded in perturbation theory, one can relate the results of the perturbed system to those of the unperturbed system. $\blacksquare$

While this example is spelled out in detail for finite dimensions, the main relations hold also in the infinite dimensional case, under certain conditions \cite{BR2}. To a state $\omega$ we associate an Hilbert space $\cH_{\omega}$ and a cyclic and separating vector $\Omega_\omega$ via the GNS construction. The modular operator associated to $(\cH_{\omega},\Omega_\omega)$ is $\Delta = \e^{-H}$ and governs the modular time evolution
\be
\alpha (\sfa) = \Delta^{-\ii t} \, \sfa \, \Delta^{\ii t}
\ee
of the operators in the algebra. If we now introduce a perturbation $V$ the system will settle into a KMS state $\omega_V$, associated with a new vector $\Omega_{\omega_V}$ via the GNS construction. The modular group automorphism is now implemented by the perturbed modular operator
\be \label{modularV}
\Delta_V = \e^{-\left(H +V - J V J\right)} 
\ee
and has the form
\be \label{autoV}
\alpha_V (\sfa) = \Delta_V^{-\ii t} \, \sfa \, \Delta_V^{\ii t} \, .
\ee
Note that the conjugate term $J V J$ is necessary to ensure that 
\be \label{commutewithJ}
\left[ H +V - J V J , J \right] =  \left[ V - J V J , J \right] = 0
\ee
in the case when $\left[ V, J \right] \neq 0$. In the literature the operators $H$ and $H +V - J V J$ are often called the standard Liouvilleans $L$ and $L_V$ associated with the evolution of the system.

When rephrased in terms of the standard Liouvillean the study of NESS becomes a spectral problem \cite{JP1}. Namely a NESS $\eta$ which is in the kernel of $L_V$ is a KMS state for which $\Delta_V$ is the modular operator. On the other hand if $\ker L_V = \{ \emptyset \}$ any NESS in $\Sigma_V^{\pm} (\omega)$ is a genuinely nonequilibrium state. Note that the operator \eqref{modularV} and the automorphism \eqref{autoV} both make sense even if the system does not settle into a KMS state $\omega_V$ but into a more general NESS, which needs not be modular. This is similar to ordinary quantum field theory, where a Wick rotation relates the evolution operator to the thermal equilibrium state; however the former makes sense even when system is not in thermal equilibrium.

Now suppose we find ourselves in the following situation, which we will encounter when studying gravitational algebras. We have a KMS state and its modular operator. Now we perturb the system and let it evolve into a NESS which is not in thermal equilibrium and in particular needs not to be modular, but for some reason we don't know how to write the automorphism $\alpha_V$, maybe because the algebra is too complicated. A solution present itself if we pretend that the system settles into a KMS state and we are able compute its modular operator. In this case from the modular operator we can read the standard Liouvillean and compute the automorphism $\alpha_V$. This is just a version of a familiar statement in statistical mechanics: the condition of thermal equilibrium requires that the statistical weight of a state is determined by the generator of the dynamics.

Going back to the finite dimensional example outlined above, one starts with a KMS state and ends up with another KMS state after introducing a local time independent perturbation. In this case, as we will see, the entropy production of the perturbed state $\omega_V$ is zero. Note that, while in this case the thermodynamics behaviour does not give anything new, there are still interesting observables one can study.

On the other hand the simplest way to engineer a nontrivial NESS is via coupling to a set of external reservoirs. By choosing appropriately the temperatures of the reservoirs one can insure that the initial state of the system is not in thermal equilibrium (but eventually a product of KMS states at different temperatures). We will see in the next Section how to study the entropy production in such a state.

\subsection{Adding reservoirs}

In quantum mechanics a typical construction of a NESS consists in the coupling of a finite dimensional quantum system $S$ to a collection of reservoirs. The original system is the quantum dynamical system $(\cA_S , \alpha_S)$. Also the reservoirs are described by an operator algebra.

More precisely we assume for the reservoirs the following structure. A collection of reservoirs is modelled by a certain dynamical system $(\cO_\cR , \alpha_\cR)$, where the dynamics is generated by $\delta_\cR$. The reservoir $\cR$ consists of $M$ parts $\cR_1 , \dots , \cR_M$, each with its own dynamics $\alpha_{\cR_i}$ generated by $\delta_{\cR_i}$, so that $\alpha_\cR = \otimes_{j=1}^M \alpha_{\cR_j}$. Furthermore we will assume that each reservoir is in thermal equilibrium and described by a $\alpha_{\cR_i}$-KMS state $\omega_{i}$ at a certain inverse temperature $\beta_i$.

The joint system is described by the algebra $\cA = \cA_S \otimes \cO_{\cR_1} \otimes \cdots \otimes \cO_{R_M}$. Sometimes we will denote by $\cO_b = \bigotimes_{i=1}^M \cO_{\cR_i}$ the operator algebra associated with the reservoir. We assume the joint system to be initially decoupled and described by a quantum dynamics $(\cA , \alpha)$ with $\alpha = \otimes_{a=0}^M \, \alpha_a = \alpha_S \otimes \alpha_\cR$.

A clarification on the nomenclature. A single state $\omega_k$ as above is a $(\alpha_{\cR_k}^t , \beta_k)$-KMS state. By rescaling this means that it is also a $(\alpha_{\cR_k}^{-t/\beta_k} , -1)$-KMS state. Therefore if we have a product state $\omega = \otimes_k \, \omega_{\cR_k}$ where each factor is  $(\alpha_{\cR_k}^t , \beta_k)$-KMS, this implies that $\omega$ is $(\otimes_k \alpha_{\cR_k}^{-t/\beta_k} , -1)$-KMS. However this does \textit{not} imply that when we put the reservoirs in contact with the system the overall system is in thermal equilibrium. This is certainly not the case unless all the temperatures are all equal. The relation between thermal equilibrium and the KMS condition holds when a state is KMS with respect to the physical Hamiltonian evolution, and not with respect to the rescaled products $\otimes_k \alpha_{\cR_k}^{-t/\beta_k}$.

The original system and the reservoirs are coupled via a perturbation $V = \sum_{j=1}^M V_j$ where each $V_j = V_j^\dagger \in \cA_0 \otimes \cO_j$ models the interaction between $\cA_S$ and $\cO_{\cR_j}$. The perturbation induces an automorphism $\alpha_V$ of $\cA$. Note that now operators in the system $S$, while they still commute with every operator in the reservoirs, do not necessarily commute with the interaction.

Now let us turn to the case of interest, that of a gravitational algebra which arises from a crossed product.

We couple the left and right boundaries with two infinite reservoirs, eventually partitioned in sub-systems. The reservoirs will be now described by two operator algebras $\cO_{b,R}$ and $\cO_{b,L}$. We can for example assume that the reservoirs are in equilibrium (multi)-Gibbs states with Hamiltonians $H_{\omega,R}$ and $H_{\omega,L}$ (in the same notation as above, except for the left $L$ and right $R$ specifiers). For simplicity we assume that these hamiltonians make sense in the large N limit\footnote{Most of the times we will not need this assumption. We will however do in a situation where we will consider the system only coupled to reservoirs on the right side. We therefore assume that we have no problem in defining traces and density matrices in the reservoirs, decoupled from the boundary theory.}. Nevertheless it is convenient to associate a vector $\Omega_\omega$ in a Hilbert space to the functional $\omega$. Since $\omega$ is the functional which gives the vacuum expectation value of operators in the thermodynamic limit of a Gibbs state, its GNS construction is essentially the thermofield double construction \cite{Witten:2021jzq}. Therefore we can take the GNS vector $\Omega_\omega$ as a thermofield double state associated to the reservoirs. In particular in the $i^{\rm th}$ partition of the reservoir the vector $\ket{\Omega_{\omega_i}}$ is annihilated by the Hamiltonian $\widehat{H}_{\omega_i} = H_{\omega_i , R} - H_{\omega_i , L}$. The modular Hamiltonian is then $\Delta_{\omega_i} = \e^{- \beta_i \widehat{H}_{\omega_i}}$.

In the following we will use a compact notation where expression of the form $\beta \widehat{H}_\omega$ really mean $\sum_{j=1}^M \beta_i \widehat{H}_{\omega_i}$ leaving open the possibility that the reservoir $\cR$ is partitioned in $M$ components at different inverse temperatures $\beta_i$. 

If for example one considers the right algebra then
\be
\cA_{0,R} \otimes \cO_{b,R}
\ee
is the total algebra at $N = \infty$ in the right exterior of the black hole. At the moment the two algebras are still decoupled. Adding the central element (which in the bulk is related to the black hole mass) gives
\be
\cA_{0,R} \otimes \cA_{U} \otimes \cO_{b,R}
\ee
where however the two algebras are still decoupled. Note that $U$ commutes with all of $\cA_{0,R}$ and obviously also with every operator in $\cO_{b,R}$.

If one wants to include perturbative $1 / N$ corrections, since the algebras are still decoupled we can simply proceed as in Section \ref{GravAlg}. The decoupled algebra $\cA_{0,R} \otimes \cO_{b,R}$ has the diagonal automorphism
\be
(\Delta_\Psi \otimes \Delta_\omega)^{-\ii s} \left( \cA_{0,R} \otimes \cO_{b,R} \right) (\Delta_\Psi \otimes \Delta_\omega)^{\ii s}
\ee
To this automorphism one associates the dynamics of the system
\be
\tau^s (\sfa) = \e^{\ii \left(\widehat{H} + \widehat{H}_\omega \right) s} \sfa  \e^{- \ii \left(\widehat{H} + \widehat{H}_\omega \right) s} 
\ee
and its generator $\delta$. Note that $\delta$ generates the flow both of the system and of the reservoir. We will sometimes drop the $s$ from $\tau^s$.

The automorphism above is not inner because of the factor $\Delta_\Psi$ (since we have assumed the bath Hamiltonians exist and are part of the bath algebras, and therefore so it is $ \Delta_\omega$). To include it in the algebra one needs to take the crossed product. The bath algebra commutes with the large N algebra and therefore does not play any role in the crossed product. The result is therefore the algebra
\be
\cA^{(b)}_R = \cA_{0,R} \rtimes \real_\Psi \otimes \cO_{b,R} \, .
\ee
This is the product of two decoupled algebras. It is however worth spelling out some of its structure in more detail. 

Operators in this algebra have the form $\sfa \e^{- \ii s \widehat{h}}  \otimes \e^{\ii s X} \otimes o_b$, acting on the Hilbert space $\cH \otimes L^2 (\real) \otimes \cH_{\omega}$. 

The algebra $\cA^{(b)}_R$ posses a special class of states, the classical-quantum states, of the form $\Psi \otimes g (X) \otimes \Omega_\omega$. Similarly a trace can be defined using the product structure, by taking the trace separately over the two decoupled factors, where the trace over the $\cA_{0,R} \rtimes \real_\Psi$ factor was defined in \eqref{traceII}. We assume that the bath is accurately described by Gibbs states, which therefore have well defined Hamiltonians and traces. Or we could assume that the bath is a copy of the boundary theory in its ground state.

Now the modular operator is again a tensor product, since the two systems are still decoupled, but the first factor is replaced by $\widehat{\Delta}_\Psi = K \widetilde{K}$. Now we have the modular flow given by
\be
K^{-\ii s} \, \sfa \, K^{ \ii s} \otimes \Delta_\omega^{-\ii s} \, o_b \, \Delta_\omega^{\ii s} = \left( K \otimes \Delta_\omega \right)^{-\ii s} \sfa \otimes o_b \left( K \otimes \Delta_\omega \right)^{\ii s}
\ee
and consequently the dynamics is generated by
\be \label{WHtau}
\widehat{\tau}^s (\sfa \otimes o_b) = \e^{\ii s \left( I + \widehat{H}_\omega \right)}  \left( \sfa \otimes o_b \right) 
 \e^{- \ii s \left( I + \widehat{H}_\omega \right)} 
\ee
in terms of the modular hamiltonian
\be \label{modH-I}
I = \widehat{H} + \frac{X}{\beta} - \frac{1}{\beta} \log g \left( \beta \widehat{H} + X \right) \, .
\ee
We will loosely refer to \eqref{WHtau} and its generalizations as ``dynamics'' and to the evolution parameter as ``time'' even if technically we are talking about a modular flow. The reason for this is that it is the modular flow which descends directly from the physical dynamics of the system after including perturbative $1 / N$ corrections. It contains the physical Hamiltonian responsible for time evolution, as well as other corrections which are necessary for a proper algebraic treatment.

Next we turn to the study of the coupled algebra $\cA^{(b)}_{R,V}$

\subsection{Interacting algebras and perturbed KMS states} \label{IntAlgandKMS}

Now we introduce an interaction term in the total Hamiltonian and couple the two algebras. We want to understand the form of the Liouvillean / modular Hamiltonian for the coupled algebras at $N=\infty$ and subsequently to include $1/N$ corrections. Note however that the would be modular operator is not anymore associated with a tensor product state such as $\Psi \otimes \Omega_\omega$.

To couple the two algebras we introduce an interaction so that the total Hamiltonians have the structure
\begin{align} \label{Htot}
H_{\mathrm{tot},R} = H_R + H_{\omega,R} + V_R \\ \nonumber
H_{\mathrm{tot},L} = H_L + H_{\omega,L} + V_L
\end{align}
where $V_R$ and $V_L$ are self-adjoint operators which mediate the interaction between the original systems and the reservoirs and are conjugate to each other by the operator $J$. For example
\be
V_R = \sum_{j=1}^M V_{R,j}
\ee
where $V_{R,j}j \in \cA_{R} \otimes \cO_{b_j,R}$. We couple the boundary theory and the reservoirs in this way to keep the formalism as general as possible. A particular case of this coupling, which we will consider in an example, is when the left hand side Hamiltonian and interactions are trivial and the system is coupled to a reservoir only on the right side (assuming now $V_R$ and $J$ commute). We will use the notation $\cA_{0,R} \otimes_V \cO_{b,R}$ to remind us that the algebra of operators now contains interactions.

Note however that the Hamiltonians \eqref{Htot} do not exist in the large $N$ limit, since $H_R$ and $H_L$ do not exist separately. We however have assumed that the reservoirs' Hamiltonians have a well defined $N$ limit (for example do not depend on $N$ altogether). The reservoir on the left boundary are conjugated to those on the right boundary; for example the left reservoir algebra is understood to be the commutant of the right reservoir algebra, to couple consistently with the boundary theories.

As we are assuming the reservoir to be less intrusive as possible with respect to the dynamics of the bulk, we can still assume that the product $\cA_{0,R} \otimes_V \cO_{b,R}$ is a Type $\mathrm{III}_1$ algebra. Taking the crossed product we will therefore obtain a Type $\mathrm{II}_\infty$ algebra, as with the $\cA_{0,R}$ factor alone.

However in general the perturbed state will not be the a tensor product involving the thermofield double state $\Psi$ anymore. There are a few alternatives to treat the problem. We could provide some educated guess for the state which replaces $\Psi$, for example another KMS state or a genuinely nonequilibrium steady state. We could work in perturbation theory and study the properties of NESS perturbatively in the coupling to the bath. Or we could try to understand aspects of the system which do not require the specification of such a state; we will study one such observable associated with the entropy production in the next Section. For the time being we will investigate the first two alternatives.

We can make progress if we assume that the perturbed state is KMS. This is for example the case if all the reservoirs are at the same temperature. Consider first the decoupled theory at $N = \infty$. Then, as discussed previously, the reference state in the GNS construction is given by $\Psi \otimes \Omega_\omega$ and the modular operator is the tensor product $\Delta_{\Psi} \otimes \Delta_{\omega} = \Delta_{\Psi \otimes \omega}$. Note that $(\widehat{H} + \widehat{H}_\omega) \Psi \otimes \Omega_\omega = 0$. Let us now add a perturbation $V = V^\dagger \in \cA_{0,R} \otimes \cO_b$ and assume that the perturbed state $\Psi_V$ is $(\tau_V , \beta)$-KMS, where $\tau_V$ denotes the perturbed dynamics. This assumption is realistic for sufficiently small perturbations. From the properties of the perturbed equilibrium KMS state we will be able to write down $\tau_V$ explicitly.

Let us define
\be
\Psi_V = \e^{-\beta (\widehat{H} + \widehat{H}_\omega + V)/2} \, \Psi \otimes \Omega_\omega \, .
\ee
Note that this is not a product state. Since $J V J$ belongs to the commutant algebra, we can write
\be
\e^{\ii \left(\widehat{H} + \widehat{H}_\omega + V \right)} J V J \e^{-\ii \left( \widehat{H} + \widehat{H}_\omega + V \right)} = \e^{\ii \left( \widehat{H} + \widehat{H}_\omega \right)} J V J \e^{-\ii \left( \widehat{H} + \widehat{H}_\omega \right)} 
\ee
Then
\begin{align}
& \e^{\ii (\widehat{H} + \widehat{H}_\omega + V) t} \left(\widehat{H} + \widehat{H}_\omega + V - JVJ \right) \e^{-\ii (\widehat{H} + \widehat{H}_\omega + V) t} \left( \Psi \otimes \Omega_\omega \right) 
\\ \nonumber & = \left[ \widehat{H} + \widehat{H}_\omega + V  - \e^{\ii (\widehat{H} + \widehat{H}_\omega) t} \left( JVJ \right) \e^{-\ii (\widehat{H} + \widehat{H}_\omega ) t} \right] \left( \Psi \otimes \Omega_\omega \right) 
\\ \nonumber & = \left( V - \e^{\ii (\widehat{H} + \widehat{H}_\omega ) t} J V \right) \left( \Psi \otimes \Omega_\omega \right) 
\end{align}
where we have used $(\widehat{H} + \widehat{H}_\omega) \left( \Psi \otimes \Omega_\omega \right) = 0$. Note that $\e^{\ii (\widehat{H} + \widehat{H}_\omega) t}  = \Delta_{\Psi \otimes \omega}^{- \ii t/\beta}$. Now we use the KMS assumption to analytically continue the last expression to $t = - \ii \beta/2$ so that
\begin{align}
\left( V - \Delta_{\Psi \otimes \omega}^{-1/2} J V \right) \left( \Psi \otimes \Omega_\omega \right) =  \left( V - J \Delta_{\Psi \otimes \omega}^{1/2} V \right) \left( \Psi \otimes \Omega_\omega \right) = \left( V - V^\dagger \right)  \left( \Psi \otimes \Omega_\omega \right) = 0
\end{align}
Retracing back our steps, we have shown that
\be
 \left( \widehat{H} + \widehat{H}_\omega+ V - JVJ \right) \e^{-\ii (\widehat{H} + \widehat{H}_\omega + V) (-\ii \beta/2)} \left( \Psi \otimes \Omega_\omega \right) = 0
\ee
or equivalently that 
\be
\Psi_V = \e^{-\beta (\widehat{H} + \widehat{H}_\omega + V)/2} \left( \Psi \otimes \Omega_\omega \right)
\ee
is annihilated by the operator $L_V =  \widehat{H} + \widehat{H}_\omega + V - JVJ$. In other words the KMS condition forces $L_V$ to have a zero eigenvalues. This is in accordance with the discussion in the paragraph below equation \eqref{commutewithJ}. In particular the state $\Psi_V$ is modular (cyclic and separating) and its modular operator is 
\be
\Delta_{\Psi_V} = \e^{- \beta L_V} = \e^{-  \beta \left( \widehat{H} + \widehat{H}_\omega+ V - JVJ \right)}\, .
\ee
Note that in the whole derivation it is crucial that all the subsystems are at the same inverse temperature $\beta$. 

To uniformize with the notation of \eqref{Htot}, recall that $V = V_R$ and $J  V J = V_L$. We set now $\widehat{V} = V_R - V_L$ so that the Liouvillean has the simple form $L_V =  \widehat{H} + \widehat{H}_\omega + \widehat{V}$.

Note that from the explicit form of $\Delta_{\Psi_V}$ we can determine the dynamics $\tau_V$ of the system
\be
\tau_V^s \left( \sfa_V \right) = \e^{\ii s L_V} \sfa_V \e^{- \ii s L_V} 
\ee
for any $\sfa_V \in \cA_{0,R} \otimes_V \cO_{b,R}$. Note that this is precisely the evolution operator we could have guessed from the general form of the Hamiltonians \eqref{Htot}. Determining the evolution operator will become less trivial once we include perturbative $1 / N$ corrections.

Now to incorporate $1/N$ corrections we take the crossed product. The construction goes on as before with minor modifications.
With the same notation of Section \ref{GravAlg}, in order to carry on the crossed product in the coupled algebras we now set 
\be
T = \beta \left( \widehat{H} + \widehat{H}_\omega+ \widehat{V}  \right) = \beta L_V \ee
and 
\be
X = \beta N U_L
\ee
Now $\e^{\ii T s}$ is an outer automorphism for the tensor product algebra $\cA_{0,R} \otimes_V \cO_{b,R}$ (the notation $\otimes_V$ is just a reminder that now the two algebras are interacting). The reason for this is that it is an outer automorphism for the algebra $\cA_{0,R}$ and that the commutators $[\widehat{H} , V]$ and $[\widehat{H}_{\omega} , V]$ will in general be  both non trivial. Taking the crossed product results now in the algebra $\cA^{(b)}_{R,V} = \left( \cA_{0,R} \otimes_V \cO_{b,R} \right) \rtimes \real_{\Psi}$.

Before we proceed a brief remark about the crossed product. We have chosen the operator $X$ to contain only information about the boundary theory and not the reservoir. We could have also involved the reservoirs more directly in the crossed product by adjoining bounded functions which depend also on $H_{L,\omega}$. The fact that we didn't correspond physically to the assumption that the reservoir does not gravitate and is not affected by $1 / N$ corrections. We stress that this could be changed if needed. 

Going back to the algebra $\cA^{(b)}_V$ we are now interested in studying some of its properties. Let us begin with the modular operator. We make the following ansatz
\be
\widehat{\Delta}_{\Psi_V} = \Delta_{\Psi_V}  g \left( \beta \left( \widehat{H} + \widehat{H}_\omega+ \widehat{V} \right)+ X \right) \, g (X)^{-1} =  \Delta_{\Psi_V}  g \left(\beta \, L_V + X \right) \, g (X)^{-1}
\ee
and take
\be
\widehat{\Psi}_V = \Psi_V \otimes g(X)^{1/2}
\ee
as the classical-quantum state of interest.

Let us check these statements. As in Section \ref{GravAlg} the algebra is generated additively by operators of the form $\widehat{\sfa}_V = \sfa_V \e^{\ii u \left( \beta \, L_V + X \right)}$ and it is therefore enough to check these statements on these operators. Note that in general the operator $\sfa_V$ is not in a factorized form. This can be easily see starting from a factorized operator in the $N= \infty$ theory and then applying the interacting time evolution operator; the result of this operation will in general not be factorizable.

To begin with we must have
\be
\braket{\widehat{\Psi}_V \vert \widehat{\sfa}_V \widehat{\sfb}_V \vert \widehat{\Psi}_V}
= \braket{\widehat{\Psi}_V \vert \widehat{\sfb}_V \widehat{\Delta}_{\Psi_V} \widehat{\sfa}_V \vert \widehat{\Psi}_V}
\ee
by definition of the modular operator. Indeed one can follow almost verbatim \cite{Witten:2021unn} and write
\be
\widehat{\Delta}_{\Psi_V} = \frac{\Delta_{\Psi_V}}{2 \pi g (X)} \int_{-\infty}^{+\infty} \dd w  \e^{- \ii w \left( \beta \,  L_V + X \right)}  \tilde{g} (w)
\ee
where $\tilde{g} (w)$ is the Fourier transform of $g (X)$. Then we see that
\begin{align} \label{modularLHS}
\braket{\widehat{\Psi}_V \vert \widehat{\sfa}_V \widehat{\sfb}_V \vert \widehat{\Psi}_V} & = \int_{-\infty}^{+\infty} \dd X g(X) \braket{\Psi_V \vert  \sfa_V \e^{\ii s \left( \beta  L_V + X \right)} \,  \sfb_V \e^{\ii s \left(  \beta L_V + X \right)} \vert \Psi_V} \\
& = \int_{-\infty}^{+\infty} \dd X g(X) \e^{\ii s X} \e^{\ii t X} \braket{\Psi_V \vert  \sfa_V \e^{\ii s \beta  L_V} \,  \sfb_V \e^{-\ii s  \beta L_V} \e^{\ii s  \beta  L_V} \e^{\ii t  \beta L_V} \vert \Psi_V} \cr \nonumber
& = \tilde{g} (s+t) \braket{\Psi_V \vert  \sfa_V \e^{\ii s \beta L_V} \,  \sfb_V \e^{-\ii s \beta L_V} \vert \Psi_V}
\end{align}
after performing the integration over $X$. We have used repeatedly the fact that $L_V \Psi_V = 0$, which follows from the fact that $\Psi_V$ is a KMS state and $\Delta_{\Psi_V}$ its modular operator. This could be no longer true if $\Psi_V$ were a genuinely nonequilibrium state.

On the other hand we can write
\begin{align}
& \braket{\widehat{\Psi}_V \vert \widehat{\sfb}_V \widehat{\Delta}_{\Psi_V} \widehat{\sfa}_V \vert \widehat{\Psi}_V}  \\ & = \int_{-\infty}^{+ \infty} \dd X
 \braket{\Psi_V \vert  \sfb_V \e^{\ii t \left( \beta L_V + X \right)} \frac{\Delta_{\Psi_V}}{2 \pi} \int_{-\infty}^{+\infty} \dd w \, \e^{- \ii w (\beta L_V + X )} \tilde{g} ( w)    \sfa_V \e^{\ii s \left( \beta L_V + X \right)} \vert \Psi_V} 
 \cr \nonumber & =
 \int_{-\infty}^{+\infty} \dd w \frac{\tilde{g} (w)}{2 \pi} \int_{-\infty}^{+ \infty} \dd X \e^{\ii (t-w+s) X}  
  \braket{\Psi_V \vert  \sfb_V \e^{\ii t \beta L_V } \Delta_{\Psi_V}  \, \e^{- \ii w \beta L_V } \tilde{g} ( w)    \sfa_V \e^{\ii s \beta L_V} \vert \Psi_V} 
\cr \nonumber
& = \int_{-\infty}^{+\infty} \dd w \frac{\tilde{g} (w)}{2 \pi} \int_{-\infty}^{+ \infty} \dd X \e^{\ii (t-w+s) X}  
 \braket{\Psi_V \vert  \sfb_V \e^{\ii (t + \ii - w) \beta L_V }  \sfa_V \e^{-\ii (t + \ii - w) \beta L_V} \vert \Psi_V} 
 \cr \nonumber
 & = \tilde{g} (t+s)  \braket{\Psi_V \vert  \sfb_V \e^{\ii (\ii - s) \beta L_V }  \sfa_V \e^{-\ii (\ii - s) \beta L_V} \vert \Psi_V} 
\end{align}
and the equality with \eqref{modularLHS} follows from the KMS condition and time translation symmetry.

Finally as in Section \ref{GravAlg} the modular operator factorizes as
\be
\widehat{\Delta}_{\Psi_V} = \widetilde{\cK}_V \, \cK_V
\ee
where
\begin{align}
\cK_V & = \e^{- \left(\beta L_V + X \right) } g \left( \beta L_V + X \right) = \e^{- \beta \left[ L_V + \frac{X}{\beta} - \frac{1}{\beta} \log g \left( \beta L_V + X \right) \right]}\cr 
\widetilde{\cK}_V & = \frac{\e^X}{g (X)} \, .
\end{align}
In particular note that this factorization implies that the $*$-automorphism
\be \label{tauVhat}
\widehat{\tau}_V^s \left( \widehat{\sfa}_V \right) = \widehat{\Delta}_{\Psi_V}^{- \ii s/\beta} \, \widehat{\sfa}_V \, \widehat{\Delta}_{\Psi_V}^{\ii s/ \beta} = \cK_V^{- \ii s / \beta} \, \widehat{\sfa}_V \, \cK_V^{\ii s / \beta}
\ee
is now inner, since $\cK_V$ is now an element of the crossed product algebra $\cA^{(b)}_{R,V}$. We interpret the automorphism $\widehat{\tau}_V$ as the natural modular time evolution of the quantum dynamical system describing the gravitational algebras coupled to reservoirs, once $1/ N$ corrections have been accounted for. It can be described as the  conjugation 
\be
\widehat{\tau}_V^s \left( \widehat{\sfa}_V \right) = \e^{\ii t I_V} \,  \widehat{\sfa}_V \,  \e^{-\ii t I_V} 
\ee
in terms of the modular Hamiltonian
\be \label{modH-IV}
I_V = L_V + \frac{X}{\beta} - \frac{1}{\beta} \log g \left( \beta L_V + X \right) \, .
\ee
To clarify our approach, we initiate our analysis from the thermofield double state, which characterizes thermal physics in equilibrium. The introduction of perturbative $1/N$ corrections leads to the emergence of the automorphism induced by $K$  as described in Section \ref{GravAlg}. This automorphism notably is associated to an equilibrium state. The modular hamiltonian contains the exponential of $\beta \widehat{H} + X$, which is a conceptual analogue to the non-existent (in the large $N$ limit) operator $H_R$ governing time evolution on the right boundary. The supplementary term $\log g (\beta \widehat{H} + X)$ accounts for gravitational corrections. Subsequently, we couple the system to reservoirs at the same temperature, ensuring that the resultant state is again at thermal equilibrium. Employing Araki's general theory of perturbation of KMS structures, we can explicitly express and compute the modular hamiltonian for this state. Furthermore Araki's theory assures us that the deformed KMS state is unique \cite{BR2}. This state corresponds to an equilibrium state and its modular hamiltonian is naturally identified with the physical dynamics, taking into account gravitational corrections. We therefore interpret $I_V$ as the generator of the dynamics of the system and we will use it to study the system even out of equilibrium. Note that in general any modular state is KMS with respect to its modular operator; in our case the modular state descends directly from the thermofield double and therefore we can interpret its modular operator as the generator of the physical dynamics.  This interpretation mirrors the fundamental relationship between equilibrium statistical weights and the evolution operator in statistical mechanics.

We can now define a trace as in Section \ref{GravAlg}
\be
\tr \,  \widehat{\sfa}_V =\braket{\widehat{\Psi}_V \vert \widehat{\sfa}_V \cK_V^{-1} \vert \widehat{\Psi}_V}   = \braket{\widehat{\Psi}_V \vert \widehat{\sfa}_V \frac{\e^X}{g (X)} \vert \widehat{\Psi}_V} = \int_{-\infty}^{+ \infty} \dd X \, \e^X 
\braket{\Psi_V \vert \widehat{\sfa}_V  \vert \Psi_V}
\ee
which is finite for a certain subalgebra, characterized by the fact that the above integral is convergent. Note that again we have used the fact that $L_V \Psi_V = 0$. The same caveats discussed in Section \ref{GravAlg} also apply in this context. Note that this definition is identical to \eqref{traceII}, except that now the state $\Psi$ is replaced by $\Psi_V$. 

By using the definition of the trace one can also define density matrices and therefore the von Neumann entropy. In particular the density matrix corresponding to the classical-quantum state $\widehat{\Psi}_V$ is $\cK_V$ itself, since
\be \label{TrPsiV}
\Tr  \, \widehat{\sfa}_V \, \cK_V = \braket{ \widehat{\Psi}_V \vert \cK_V \cK_V^{-1}   \widehat{\sfa}_V  \vert \widehat{\Psi}_V } = \braket{ \widehat{\Psi}_V \vert  \widehat{\sfa}_V  \vert \widehat{\Psi}_V } \, .
\ee

So far our analysis has followed  \cite{Witten:2021unn} closely and the generalization of every proposition there to our case was almost verbatim. However now we come to a very important physical point. Since
\be
\cK_V \, \log \cK_V = \e^{- \left( \beta L_V + X \right)} g \left( \beta L_V + X \right) \left(
- \left( \beta L_V + X \right) + \log g \left( \beta L_V + X \right)
\right)
\ee
by using the definition of the trace \eqref{TrPsiV} we see that the von Neumann entropy for the state $\widehat{\Psi}_V$ is
\be \label{vnPsiV}
S (\widehat{\Psi}_V)_{\cA^{(b)}_{R,V}}  = \int_{-\infty}^{+ \infty} \dd X \left( X g(X) - g(X) \log g(X) \right)
\ee
precisely as in \eqref{vnPsi}! The entropy for the weakly interacting theory in thermal equilibrium is the same as the entropy for the non-interacting theory, which up to a constant coincides with the entropy of the isolated system without the reservoirs (because of thermal equilibrium). In other words we have proven explicitly that a weak time-independent interaction which perturbs the original state into a new thermal state does not have any non-trivial thermodynamics and in particular does not give rise to entropy production. This is the counterpart of the familiar result in thermodynamics where adiabatic processes do not involve entropy production and it is a consequence of the fact that as the dynamics is perturbed by the interaction, also the ground state changes accordingly. We will find again this result in the next Section from a more formal perspective. In this construction it was crucial to assume that the reservoirs and the bulk are all at the same temperature. The situation will change drastically when we will allow the reservoirs to have different equilibrium temperature and/or the perturbation to be time dependent, which is the case of physical interest.

\subsection{Interacting algebras and general NESS}

In the general case the NESS of the interacting algebra will not be a KMS state. Nevertheless modular time evolution in the interacting algebra $\cA^{(b)}_{R,V}$ will still be given by the automorphism $\widehat{\tau}_V$ defined in \eqref{tauVhat}. This is just an example of the general relation between the statistical weight of a thermal state and the evolution operator. By studying the KMS states of the algebra $\cA^{(b)}_{R,V}$ we have found their statistical weights from the modular operator and therefore the $*$-automorphism which determined modular time evolution in the algebra. We can now use it without any reference to thermal equilibrium. This will allow us to characterize NESS in perturbation theory.

Now we consider a general time dependent perturbation and we assume that the system settles into a NESS which is not a KMS state. What would such a state look like? We can make some progress by adapting to the gravitational algebras the perspective advocated in \cite{RuelleRev,RuelleNESS}

Assume we start with the equilibrium classical-quantum state\footnote{We are abusing notation here since this $\widehat{\Psi}$ is not the same object we encountered in Section \ref{GravAlg}. We trust this will not cause confusion since now every vector will  involve also the reservoirs. Also the position of the $g (X)$ factor in the tensor product does not matter.} $\widehat{\Psi} = \Psi \otimes \Omega_\omega \otimes \sqrt{g (X)}$ in the noninteracting gravitational algebra $\cA^{(b)}_R$. We let evolve the system after a perturbation is applied. The above state is an equilibrium state and therefore invariant under the decoupled evolution $\widehat{\tau}$ defined in \eqref{WHtau}. If we consider a time dependent perturbation which vanishes in the far past, the state $\widehat{\Psi}$ is also an element of the algebra $\cA^{(b)}_{R,V}$.

We define the \textit{time dependent} state
\be \label{NNES}
\chi_t (\widehat{\sfa}_V) =  \lim_{s \rightarrow - \infty} \braket{ \widehat{\Psi} \vert \widehat{\tau}_V (s , t)\left( \widehat{\sfa}_V \right) \vert \widehat{\Psi}} \, .
\ee
In plain words $\chi_t$ is a state reached at time $t$ from the evolution of the interacting system starting from the state $\widehat{\Psi}$ in the infinite past, before the perturbation was turned on. Here\footnote{In ordinary quantum mechanics this would be just the operator $\e^{-\ii (t_0 - t) H}$ responsible for the evolution of the system from time $t_0$ to time $t$.}
\be
\widehat{\tau}_V (s,t)= \widehat{\tau}_V^{\, -s} \, \widehat{\tau}_V^{ \,t}
\ee
which becomes simply $\widehat{\tau}_V^{\, t-s}$ if the perturbation is time independent or if it commutes with itself at different times. Similarly $\widehat{\tau} (s,t) = \widehat{\tau}^{\, t-s}$ for the free evolution. Since the vector $\widehat{\Psi}$ is by assumption invariant under the free evolution $\widehat{\tau}$, we can rewrite \eqref{NNES} as
\be 
\chi_t (\sfa) = \lim_{s \rightarrow - \infty} \braket{ \widehat{\Psi} \vert \widehat{\tau} (t,s) \,  \widehat{\tau}_V (s , t)  \widehat{\sfa}_V \vert \widehat{\Psi}} \, .
\ee
Given the explicit form of the evolution operators we have the operator identity
\be \label{derInt}
\frac{\dd}{\dd u} \widehat{\tau} (t,u) \, \widehat{\tau}_V (u,t) \widehat{\sfa}_V = - \ii \widehat{\tau} (t,u) \, \left[ \cV (u) , \widehat{\tau}_V (u,t) \widehat{\sfa}_V \right] \, ,
\ee
where we have introduced the operator
\be \label{IntPart}
\cV (u) = I_V (u) - I - \sum_{j=1}^M \widehat{H}_{\omega_j} \, ,
\ee
which captures the interacting part of the modular evolution. Note that $\cV$ depends explicitly on time if the interaction $V$ depends explicitly on time.

Let us derive this expression explicitly. To begin with we introduce an analog of the interaction representation by writing
\be 
\widehat{\tau}_V^{\, t} (\widehat{\sfa}_V) = \Gamma_V^t \tau^t (\widehat{\sfa}_V) \Gamma_V^{t \, \dagger} \, .
\ee
where the $\Gamma_V^t$ is a family of unitary operators which are solution of the differential equation
\be \label{IntRep}
\frac{\dd}{\dd t} \, \Gamma^t_V = \ii \Gamma^t_V \widehat{\tau}^{ \, t} \left( \cV(t) \right) 
\ee
with initial condition $\Gamma^0_V = \mathrm{id}$.

Let us begin with the simpler identity
\be \label{derInt2}
\frac{\dd}{\dd t} \widehat{\tau}^{\, t}  \left( \widehat{\tau}^{\, t}_V \right)^{-1} \widehat{\sfa}_V = - \ii \widehat{\tau}^{\, t} \left[ \cV(t) \, , \, \left( \widehat{\tau}^{\, t}_V \right)^{-1} \widehat{\sfa}_V \right] \, .
\ee
Since
\be
\left( \widehat{\tau}^{\, t}_V \right)^{-1} (\widehat{\sfa}_V) = \widehat{\tau}^{-t} \left( \Gamma_V^{t \dagger} \, \widehat{\sfa}_V \,  \Gamma_V^t \right)
\ee
we see immediately that
\be
\widehat{\tau}^{\, t} \left( \widehat{\tau}_V^{\, t} \right)^{-1}  (\widehat{\sfa}_V) = \Gamma_V^{t  \dagger} \, \widehat{\sfa}_V \, \Gamma_V^t \, .
\ee
Therefore
\be
\frac{\dd}{\dd t}  \widehat{\tau}^{\, t}  \left( \widehat{\tau}^{\, t}_V \right)^{-1} \widehat{\sfa}_V = \frac{\dd}{\dd t} \left(  \Gamma_V^{t  \dagger} \, \widehat{\sfa}_V \, \Gamma_V^t \, \right) \, ,
\ee
and now verifying  \eqref{derInt2} simply amounts to using the differential equation \eqref{IntRep} repeatedly. Now it is easy to derive \eqref{derInt} because  it differs from \eqref{derInt2} by the composition of $t$-independent operators.

Since
\be
\int_{s}^t \dd u \frac{\dd}{\dd u}  \widehat{\tau} (t,u) \, \widehat{\tau}_V (u,t) \, \widehat{\sfa}_V = \widehat{\sfa}_V -  \widehat{\tau} (t,s) \, \widehat{\tau}_V (s,t) \, \widehat{\sfa}_V 
\ee
we can use \eqref{derInt} to find
\be
 \widehat{\sfa}_V -  \widehat{\tau} (t,s) \, \widehat{\tau}_V (s,t) \, \widehat{\sfa}_V = - \ii \int_s^t \dd u \, \widehat{\tau} (t,u) \, \left[ \cV (u) , \widehat{\tau}_V (u,t) \, \widehat{\sfa}_V \right]  \, .
\ee
Now we apply to both sides the functional corresponding to the state $\widehat{\Psi}$ and take the limit $s \rightarrow - \infty$ to find
\be
\chi_t (\widehat{\sfa}_V) = \braket{ \widehat{\Psi} \vert \, \widehat{\sfa}_V \, \vert \widehat{\Psi}  }+ \ii \int_{- \infty}^t \dd u \, \braket{ 
\widehat{\Psi} \vert
\left[ \cV(u) , \widehat{\tau}_V (u,t) \ \widehat{\sfa}_V \right]
\vert
\widehat{\Psi}
}
\ee
If now we specialize to stationary states, that is we assume that $\chi_t = \chi$ is a NESS and drop the dependence on $t$ everywhere, we obtain
\be \label{NESSru}
\chi (\widehat{\sfa}_V) = \braket{ \widehat{\Psi} \vert \, \widehat{\sfa}_V \, \vert \widehat{\Psi}  }+ \ii \int_{- \infty}^0 \dd u \, \braket{ 
\widehat{\Psi} \vert
\left[ \cV(u) , \widehat{\tau}_V^{\, -u} \ \widehat{\sfa}_V \right]
\vert
\widehat{\Psi}
} \, .
\ee
Note that the above formula is expressed in terms of the interacting evolution operator. We can rephrase it in terms of the free modular evolution operator as a Dyson series 
\begin{align}
\widehat{\tau}_V^t (\widehat{\sfa}_V) = & \widehat{\tau}^{t} (\widehat{\sfa}_V) + \sum_{n\ge1} \ii^n \int_0^t \dd t_1 \int_0^{t_1} \dd t_2 \dots \\
& \cdots \int_0^{t_{n-1}} \dd t_n \left[ \widehat{\tau}^{t_n} (\cV(t_n)) , \left[ \cdots \left[ \widehat{\tau}^{t_1} \left( \cV (t_1) \right) , \widehat{\tau}^{t_1} \left( \widehat{\sfa}_V \right) \right] \right] \right] 
\end{align}
Therefore in principle given an interaction term, then the form of the NESS state can be computed explicitly, if laboriously, in perturbation theory. Note that it will not be in general associated with a simple vector since in \eqref{NNES} taking the limit will in general not commute with taking the expectation value.

\paragraph{Remark.} One issue we do not address in this paper is the fate of symmetries in nonequilibrium steady states. In general, it is possible that the nonequilibrium dynamics will break some or all of the symmetries present in the original system. Conversely, new symmetries may emerge. Unfortunately, there is no formalism available to fully address this problem, as NESS are poorly understood compared to KMS states. However, it might be possible to use holography to provide a more concrete description. Geometric techniques from the bulk could potentially be developed to characterize nonequilibrium steady states more explicitly and to ask more precise questions about symmetries. This may be achievable in simpler models, such as JT gravity. For a discussion of symmetries within the context of crossed product algebras see \cite{AliAhmad:2024eun,Fewster:2024pur}.

\section{Entropy production} \label{ep}

In general for an arbitrary interaction it will be difficult to obtain explicit expressions for the NESS. It is therefore interesting to try to circumvent the problem by asking if there are any observables which can be computed independently of the explicit form of the NESS and to what extent. In this Section we will discuss entropy production, following the formalism of \cite{RuelleRev,JP1}.

\subsection{Entropy production in finite systems} \label{phenoEP}

Let us begin with some phenomenological aspects of entropy production in open quantum systems. If we have a system $S$ coupled to several reservoirs $\cR_j$ at different temperatures $\beta_j$, we expect to see a steady heat flow through the system. In a stationary state we expect that a non zero entropy production rate in the system $S$ is determined by the entropy flux entering / leaving the system. 

Consider a finite dimensional system $S$ whose dynamics is governed by the Hamiltonian $H_S$. In a stationary state the rate of the entropy flux entering the system is determined by the energy leaving the reservoirs
\be
- \sum_k \beta_k \Phi_k \, .
\ee
If the system-reservoir dynamics is captures by the Hamiltonian 
\be \label{Hsysres}
H = H_S + V + \sum_k H_{\cR_k} \, ,
\ee 
then the energy flux is given by Heisenberg equation
\be \label{kflux}
\Phi_k = - \ii \left[ H , H_{\cR_k} \right]  = \delta_k (V)
\ee
in terms of the generator $\delta_k = \ii \left[ H_{\cR_k} , \, \cdot \, \right]$ of the dynamics of the reservoir. We assume that the perturbation $V$ is so small that it does not affect the thermal equilibrium of the reservoirs. Therefore the entropy production in a state $\mu$ is defined as
\be \label{EPdef}
\mathrm{Ep} \left( \mu \right) = \mu \left(
- \sum_{k=1}^L \beta_k \, \delta_k (V) 
\right) \, .
\ee
We have written the entropy production in a fashion which can be immediately generalized to infinite dimensional systems.

To familiarize ourselves with the definition of entropy production  \eqref{EPdef}, consider as an illustrative example of a finite dimensional system divided into subsystems \cite{RuelleEP}. Observables in the system are elements of a von Neumann algebra of Type $\mathrm{I}_\infty$, specifically the algebra of bounded operators on the Hilbert space. We assume the full Hilbert space has the form $\cH = \otimes_a \cH_a $. The full system is described by a density matrix $\rho (t)$ which will be in general time dependent. Since the complete system is isolated, the von Neumann entropy $S (\rho) =  - \Tr \rho \log \rho$  
is however time independent, in the sense that if $U(t)$ denotes the unitary evolution operator, $S ( U(t) \, \rho \, U(-t)) = S (\rho)$. If we set $\cH_{\widehat a} = \otimes_{b \neq a} \cH_b$, we can define partial density matrices $\rho_a  = \Tr_{\cH_{\widehat a}} \rho $ and their von Neumann entropies $S_a (t) = - \Tr_{\cH_a} \rho_a \log \rho_a$, which may now be time dependent.

It is therefore natural to define the entropy production rate as
\be
e = \frac{\dd}{\dd t} \left( \sum_a S_a (t) - S \left( \rho (t) \right) \right) \, ,
\ee
where the term in parenthesis is positive by the subadditivity of the entropy and physically represent the information we lose about $\rho$ when we partition the system into its several subsystems. Note that since $S (\rho (t)) $ is really time independent, the quantity in parenthesis is just the rate of change of the entropy of every subsystem. We assume that the Hamiltonian can be written as $H = \sum_a H_a + V$ where $V$ is an interaction term and $H_a$ is the tensor product of unit matrices and the Hamiltonian $h_a$ acting on the subsystem $a$.

Then at the linearized level the evolution of the density matrix in Schr\"odinger picture is given by
\be
\rho (t + \dd t) = \e^{- \ii H \dd t} \rho \e^{\ii H \dd t} = \rho(t) - \ii \dd t \left[ \sum_a H_a + V , \rho (t) \right]
\ee
and therefore for the partial density matrices
\be
\rho_a (t + \dd t) = \rho_a (t) - \ii \dd t \left[ h_a , \rho_a \right] - \ii \dd t \, \Tr_{\cH_{\widehat a}} \left[ V , \rho (t) \right] \, .
\ee
A simple computation, using the cyclicity of the trace in each subspace, shows that
\be
S_a (t + \dd t) - S_a (t) = - \ii \dd t \, \Tr_{\cH} \left( \rho (t) \left[ V , 1_{\widehat a} \otimes \log \rho_a (t) \right] \right)
\ee
where $1_{\widehat a}$ is the identity operator on the tensor product $\cH_{\widehat a}$.

Therefore the entropy production is given by
\be \label{EPfinitedim}
e = - \ii \sum_a \Tr_{\cH} \left( \rho (t) \left[ V , 1_{\widehat a} \otimes \log \rho_a (t) \right] \right)  = - \ii \Tr_\cH \left( \rho (t)  \left[ V , \log \otimes_{a \ge 0} \rho_a \right] \right)
\ee
This formula gives a clear physical picture of the entropy production in a simple quantum mechanics system. We will now argue how the infinite volume limit reproduces \eqref{EPdef}.

To take the infinite volume limit, we declare that the subsystems with $a \neq 0$ are reservoirs in thermal equilibrium at inverse temperature $\beta_a$. Therefore we can replace their density matrices $\rho_a$ with Gibbs ensembles determined by the reservoir Hamiltonians $H_{\cR_a}$. We also assume that in the large volume limit the density matrix $\rho (t)$ of the system tends to a time independent state $\rho$. Then we can write \eqref{EPfinitedim} as
\be
e = \rho \left( \sum_a \beta_a \ii \left[ H , H_{\cR_a} \right] \right) = - \rho \left( \sum_a \beta_a \, \delta_a (V) \right)
\ee
which precisely coincides with \eqref{EPdef}; in fact one can see that the $a=0$ term can be ignored since the fluxes to the system $S$ all add up to zero in a stationary state \cite{RuelleEP}. Here $H$ is the total Hamiltonian of the system \eqref{Hsysres}. Note that while these Hamiltonians will not really make sense in the infinite volume limit, their commutator does since for local interactions it will have non vanishing support only in a finite spacetime region\footnote{This is clear for example if one thinks of an infinite lattice of spins which have a nearest neighbour interaction.}. Each commutator $\left[ H, H_{\cR_a} \right]$ represents the rate of transfer of energy (with sign) to the reservoir $\cR_a$.

\subsection{Entropy production in gravitational algebras}

We are interested in developing a similar expression in the context of gravitational algebras. To do so we pick a reference state $\widehat{\Psi} = \Psi \otimes \Omega_\omega \otimes g(X)^{1/2}$ of the non-interacting system. The main idea is that the relative entropy between this reference state and another state captures the entropy production observable and motivates its definition. Let $\delta_{\widehat{\Psi}}$ be the generator of the modular group of the reference state $\widehat{\Psi} $ 
\be \label{deltapsi}
\delta_{\widehat{\Psi}} = \ii \sum_{k=1}^L \beta_k \left[ \widehat{H}_{\omega_k} , \, \cdot \, \right] + \ii \beta \left[ I , \, \cdot \, \right]
\ee
where $\widehat{\delta}_{\cR_k} = \ii \left[ \widehat{H}_{\omega_k} , \, \cdot \, \right] $ is the infinitesimal generator of the dynamics of the $k^{\rm th}$ reservoir and the operator $I$ was defined in \eqref{modH-I}. 
Note that ${\widehat{\Psi}}$ is (+1)-KMS with respect to the dynamics generated by $\delta_{\widehat{\Psi}} $. 

Next we introduce a coupling $V$ between the boundary theory and the reservoirs. In the strict $N= \infty$ limit the two algebras interact via the coupling in the Hamiltonian, in the sense that the Hamiltonian evolution  $\tau_V$ of any operator in the tensor product algebra will contain the interaction term $\widehat{V}$. As a result any operator evolved in time will not have generically a factorized form. 

What can we say about the interaction of the two algebras when including $1 / N$ corrections? Now the situation is complicated by the crossed product. The coupling of the boundary algebra with the reservoir algebras is induced by the $N = \infty$ coupling, but it is not exactly the same since the evolution operator $\widehat{\tau}_V$ is more complicated. However it is natural to define as an interaction term the difference between the infinitesimal generator $\delta_{\widehat{\tau}_V}$ of the coupled theory and the generator $\delta_{\widehat{\tau}}$ of the uncoupled theory

We can compute $\delta_{\widehat{\tau}_V}$ from \eqref{tauVhat} as
\be \label{deltatauVhat}
\delta_{\widehat{\tau}_V} = \ii \left[ I_V \, , \, \cdot \, \right]
\ee
where $I_V$ was defined in \eqref{modH-IV}. Assume for simplicity that $V$ is time independent. Then we define the interaction
\be \label{cVint}
\cV = I_V - I - \sum_{k=1}^M \widehat{H}_{\omega_k}
\ee
as in  \eqref{IntPart}. Finally we define the entropy production observable\footnote{Since we have taken ${\widehat{\Psi}}$ to be $(+1)$-KMS the sign in the definition of entropy production is the opposite of \cite{JP1} but agrees with \cite{RuelleRev}.} as
\be \label{defsigmaV}
\sigma_V = - \delta_{\widehat{\Psi}} (\cV) \, .
\ee
Then the entropy production rate of a state $\widehat{\varphi}$, interpreted as a linear functional on the algebra $\cA^{(b)}_{R,V}$  is defined as
\be
\mathrm{Ep} (\widehat{\varphi}) =\widehat{\varphi} \left( \sigma_V \right) \, .
\ee
In particular if the state  $\widehat{\varphi}$ is a classical-quantum state represented by the vector $\widehat{\Phi}$ we simply have
\be
\mathrm{Ep} (\widehat{\varphi}) = \widehat{\varphi} (\sigma_V) = \braket{\widehat{\Phi} \, \vert \, \sigma_V \,  \vert \, \widehat{\Phi}} \, .
\ee
A more interesting case is when the state is a NESS $\chi^+ \in \Sigma^+ (\widehat{\Psi} )$ obtained starting from the (generalized) thermofield double state $\widehat{\psi}$ represented by the vector $\widehat{\Psi}$. Then
\be
\chi^+ (\sfa) = \lim_{t_k} \frac{1}{t_k} \int_0^{t_k}  \widehat{\psi} \circ \widehat{\tau}^{ \, t}_V (\sfa) \dd t =  \lim_{t_k} \frac{1}{t_k} \int_0^{t_k}  \braket{ \widehat{\Psi}  \, \vert \, \widehat{\tau}_V^{\, t} (\sfa)  \, | \, \widehat{\Psi} } \dd t \, .
\ee

What is the physical meaning of entropy production? First note that 
\be \label{sigmaV1}
\sigma_V = - \ii \sum_{k=1}^L \beta_k \, \left[ \widehat{H}_{\omega_k} , \cV \right] - \ii \beta \left[ I , \, \cV \, \right]
\ee
because of \eqref{defsigmaV} and \eqref{deltapsi}. We define 
\be \label{ThetaFlux}
\mathbf{\Theta}_k =  \delta_{\cR_k} (\cV) = \ii  \left[ \widehat{H}_{\omega_k} , \cV \right] 
\ee
which is formally the analog of \eqref{kflux} and has therefore the physical interpretation as the energy flux in/out the reservoir when $1 / N$ corrections are taken into account.

Note that by definition $\mathbf{\Theta}$ is actually the net flux entering/exiting the system from the $k^{\rm th}$ reservoir, since $\widehat{H}_{\omega_k} = H_{\omega_k , R} - H_{\omega_k , L} $. It is important to notice that, while $H_{\omega_k , R} - H_{\omega_k , L}$ gives zero on the vacuum $\Omega_{\omega_k}$, it will not be trivial on a generic state, especially after time evolution. For example a state time evolved from the reference state $\widehat{\Psi}$ will be very complicated and exhibit entanglement between the bulk and the reservoirs.

We will now connect the observable $\sigma_V$ with the phenomenological discussion in Section \ref{phenoEP}. To begin with note that $\delta_{\cR_k} (I) = 0$ since operators in the reservoir and operators in the boundary theory commute. Therefore we can write, using \eqref{deltatauVhat}
\begin{align}
\delta_{\widehat{\tau}_V} (I) & = \ii \left[ I_V , I \right] = \ii \left[ I + \sum_{k=1}^L \widehat{H}_{\omega_k} + \cV , I \right] 
= \ii \left[ \cV , I \right]
%
%
%
\end{align}
Therefore looking at \eqref{sigmaV1} we conclude that 
\be \label{sigmaV2}
\sigma_V = - \sum_{k=1}^L \beta_k \, \delta_{\cR_k} (\cV) + \delta_{\widehat{\tau}_V} (I) 
\ee
where the last term is a total derivative. We will now use this fact to show that the entropy production due to the last term is bounded and will not contribute in a $\widehat{\tau}_V$ invariant state.

Let us look now at the expression for the entropy production of a NESS $\chi^+ \in \Sigma^+ (\widehat{\Psi})$
\begin{align}
\mathrm{Ep} (\chi^+) &= \chi^+ \left( \sigma_V \right) = \lim_{t_k} \frac{1}{t_k} \int_0^{t_k}  \widehat{\psi} \circ \widehat{\tau}^{ \, t}_V \left( -  \delta_{\widehat{\Psi}} (\cV) \right) \dd t \cr
& = 
\lim_{t_k} \frac{1}{t_k}  \left[ - \sum_{j=1}^M \beta_j \int_0^{t_k}  \widehat{\psi} \circ \widehat{\tau}^{ \, t}_V \left( \mathbf{\Theta}_j \right) \dd t 
+ \int_0^{t_k}  \widehat{\psi} \circ \widehat{\tau}^{ \, t}_V \left(  \delta_{\widehat{\tau}_V} (I) \right) \dd t \right]
\cr
 & = 
 \lim_{t_k} \frac{1}{t_k}  \left[ - \sum_{j=1}^M \beta_j \int_0^{t_k}  \widehat{\psi} \circ \widehat{\tau}^{ \, t}_V \left( \mathbf{\Theta}_j \right) \dd t 
+ \widehat{\psi} \left(  \widehat{\tau}^{ \, t_k}_V (I) \right) -  \widehat{\psi} \left( I \right) \right] 
\cr
& =  \lim_{t_k} \frac{1}{t_k}  \left[ - \sum_{j=1}^M \beta_j \int_0^{t_k}  \widehat{\psi} \circ \widehat{\tau}^{ \, t}_V \left( \mathbf{\Theta}_j \right) \dd t \right] = - \sum_{j=1}^M \beta_j \, \chi^+ \left( \mathbf{\Theta}_j  \right) \, ,
\end{align}
where we have used the fact that the NESS $\chi^+$ is $\widehat{\tau}_V$ invariant to cancel the contributions coming from $\delta_{\widehat{\tau}_V} (I) $. This expression agrees with \eqref{EPdef}. Therefore our formalism incorporates gravitational corrections in such a way as to be compatible with the laws of thermodynamics.

Also note that the infinitesimal generator $\delta_{\widehat{\tau}_V} $, as well as its finite counterpart, do not reduce to the free evolution generators when the interaction is removed. This has a nice physical explanation; once the reservoirs and the bulk start interacting, even if the interactions are removed, quanta from the bulk will have passed into the reservoirs. Due to the nature of Hawking's radiation, even if the interaction is now removed, the reservoirs will still be entangled with the bulk where a black hole is present. Therefore even removing the interaction, after the systems have been put into contact, the evolution operator will not factorize into a product of the bulk and the reservoirs.

The entropy production can be written in term of the relative entropy. Again we consider the reference state $\widehat{\Psi}$ and the generator of its modular group $\delta_{\widehat{\Psi}}$. Let $U$ be a unitary operator. Then it was proven in \cite{JP2} that
\be \label{EPformulaGen}
S \left( \widehat{\Phi}^U  \Vert \widehat{\Psi} \right) =  S \left( \widehat{\Phi}  \Vert \widehat{\Psi} \right)  - \ii \widehat{\Phi} (U^\dagger \, \delta_{\widehat{\Psi}} (U))
\ee
where $\widehat{\Phi}^U $ is defined by the property that for every $\sfa \in \cA$ one has $\widehat{\Phi}^U (\sfa) = \widehat{\Phi} (U^\dagger \sfa U)$. The relation \eqref{EPformulaGen} is fairly general and requires mild regularity properties of the perturbation and the initial state $\widehat{\Psi}$ to be KMS for the dynamics induced by $\delta_{\widehat{\Psi}}$. It follows from Araki's general theory of perturbation of KMS structures.

Suppose that we have now a time dependent perturbation $V(t)$, and let $\tau_V^t$ be the perturbed evolution, as in the previous section. To deal with this case we introduce an analog of the interaction representation as in \eqref{IntRep}.  Then if we write
\be
\widehat{\tau}_V^t (\sfa) = \Gamma^t_V \, \widehat{\tau}^t (\sfa) \Gamma^{t \dagger}_V = 
 \Gamma^t_V \, \e^{\ii t (I + \widehat{H}_\omega)} \, \sfa \,  \e^{-\ii t (I + \widehat{H}_\omega)}  \Gamma^{t \dagger}_V \, ,
\ee
we see that we can include the information about the dynamics in \eqref{EPformulaGen} if we identify $U =  \e^{-\ii t (I + \widehat{H}_\omega)}  \Gamma^{t \dagger}_V$. Recall now that the initial state $\widehat{\Psi}$ is invariant under the unperturbed evolution $\widehat{\tau}$. Then for any state $\widehat{\Phi}$ if follows from \eqref{EPformulaGen} that 
\be
S \left( \widehat{\Phi} \circ \widehat{\tau}^t_V \Vert \widehat{\Psi} \right) =  S \left( \widehat{\Phi}  \Vert \widehat{\Psi} \right)  - \ii \widehat{\Phi} \left( \Gamma^t_V \, \delta_{\widehat{\Psi}} (\Gamma^{t*}_V) \right) \, .
\ee

Consider now the case of a time independent perturbation $V$, and therefore a time independent $\cV$. The following identity holds
\be \label{identityGamma}
\frac{\dd}{\dd t} \Gamma^t_\cV \,  \delta_{\widehat{\Phi}} (\Gamma^{t*}_\cV) = - \ii \tau_\cV^t \left( \delta_{\widehat{\Psi}} (\cV) \right) \, .
\ee
This can be seen by noting that for time independent interactions one has
\be
\Gamma_{\cV}^t = \e^{\ii t (I + \cV + \widehat{H}_\omega)} \, \e^{- \ii t (I + \widehat{H}_\omega)} \, ,
\ee
and therefore
\be
 \Gamma^t_\cV  \, \delta_{\widehat{\Phi}} (\Gamma^{t*}_\cV) = \e^{\ii t (I + \cV + \widehat{H}_\omega)} \, \left[
 (I + \widehat{H}_\omega) , \e^{\ii t (I + \cV + \widehat{H}_\omega)} 
 \right] \, .
\ee
One arrives to \eqref{identityGamma} by taking the time derivative of this equality. By integrating the identity \eqref{identityGamma} we find the following relation
\be \label{EntProd}
S \left(\widehat{\Phi} \circ \widehat{\tau}^t_V \Vert \widehat{\Psi} \right) =  S \left( \widehat{\Phi}  \Vert \widehat{\Psi} \right)  + \int_0^t \widehat{\Phi} \circ \widehat{\tau}_V^s (\sigma_V) \, \dd s \, ,
\ee
which expresses the entropy production in the state $\widehat{\Phi}$ in terms of the relative entropy. 

In a physically interesting situation we would choose $\widehat{\Phi}$ to be initially of the form $\Phi \otimes 1 \otimes f(X)^{1/2}$, with trivial support on the reservoir. In this case $S \left( \widehat{\Phi}  \Vert \widehat{\Psi} \right)$ can be related to the generalized entropy by using \eqref{EntropyPhi}. On the other hand $S \left(\widehat{\Phi} \circ \widehat{\tau}^t_V \Vert \widehat{\Psi} \right) $ contains a state evolved in time from $\widehat{\Phi}$ which will not have anymore trivial support on the reservoir. Indeed such a state will contain the information about the radiation which has entered the reservoir. We expect that the entropy production observable captures this information, but making this observation more precise would require new tools to deal with the time dependence in the density matrix for $\widehat{\Phi}$.

Note that the relative entropy makes sense under the assumption that $\widehat{\Phi}$ is a purely normal state, which in general will not be the case for a NESS. Still the case of a NESS can be studied as follow. Assume $\widehat{\chi}^+ \in \Sigma^+_V \left( \widehat{\Psi} \right)$ is a NESS of the perturbed system and that it can be reached by a divergent sequence $\{ \widehat{\tau}_V^{t_n} \}_{n \in \zed_+}$, then one has 
\be
\lim_{n \rightarrow \infty} \frac{1}{t_n} S \left(\widehat{\Psi} \circ \widehat{\tau}^t_V \Vert \widehat{\Psi} \right) = 
\lim_{n \rightarrow \infty} \frac{1}{t_n} \int_0^{t_n} \widehat{\Psi} \circ \tau_V^s (\sigma_V) \, \dd s = \mathrm{Ep} (\widehat{\chi}^+) 
\ee
under the assumption that the perturbation is sufficiently regular. The main point here is that, since we have related the entropy production of a NESS to the (non-decreasing) relative entropy between the reference state and its modular time-translate, we can easily conclude that
\be
\mathrm{Ep} (\widehat{\chi}^+) \ge 0
\ee
which shows that the entropy production in the NESS $\widehat{\Upsilon}$ is non-negative.

Note that the entropy production refers to the whole system, boundary theory and reservoir, and that its non-negativity is just a statement of the second law of thermodynamics. Extracting the entropy of the radiation or studying the Page curve would require more work. We will comment briefly on this in the next Section.

Let us check a few physical consequences of \eqref{EntProd}. Assume for example that the state $\widehat{\Phi}$ is a normal state which is also $\widehat{\tau}_V$ invariant. This is the case if the state $\widehat{\Phi}$ is KMS, for example of the form of the state $\widehat{\Psi}_V$ studied in the previous Section. Then \eqref{EntProd} implies that $\mathrm{Ep} \left( \widehat{\Phi} \right) = 0$, so that no nontrivial thermodynamics happens, in agreement with what we have seen in Section \ref{IntAlgandKMS}. On the other hand it may be that $\widehat{\Phi}$ is $\widehat{\tau}_V$ invariant, but not a normal state. Then one cannot use the relative entropy formula directly. We can however use the fact that normal states are dense in the weak topology to find a sequence $\widehat{\Phi}_n$ of normal states which converges to $\widehat{\Phi}$. Then we have
\be
\lim_n \left(
S \left( \widehat{\Phi}_n \circ \widehat{\tau}^t_V \vert \widehat{\Psi} \right) - S \left( \widehat{\Phi}_n  \vert \widehat{\Psi} \right)
\right)
=  \int_0^t \widehat{\Phi} \circ \widehat{\tau}_V^s (\sigma_V) \, \dd s = t \, \mathrm{Ep} (\widehat{\Phi})
\ee
where we have used the $\widehat{\tau}_V$-independence of $\widehat{\Phi}$. The latter expression has the physical interpretation as the rate of divergence of the entropy differential $\Delta S (\widehat{\Phi} , t) = t \, \mathrm{Ep} (\widehat{\Phi})$.

We can hope to make some progress in perturbation theory. To this end let us rescale the interaction term $V \longrightarrow \lambda V$, assuming for simplicity that the strength of the coupling is the same for each reservoir. The non-local structure of the modular hamiltonian complicates the perturbative expansion. For simplicity let us consider only the time independent case. For example we can expand
\be \label{expV}
\cV = \sum_{p=0}^\infty \lambda^p \, \widehat{V}^p \, c_p \left( \widehat{H} , \widehat{H}_\omega , X , \beta \right)
\ee 
where the functions $c_p$ are explicitly computable. The first terms are
\begin{align}
c_0 &= \frac{1}{\beta} \left( \log g (\beta \widehat{H} + X) - \log g (\beta \widehat{H} + \beta \widehat{H}_\omega + X) \right) \cr
c_1 &= \frac{g (\beta \widehat{H} + \beta \widehat{H}_\omega + X) - g' (\beta \widehat{H} + \beta \widehat{H}_\omega + X)}{g (\beta \widehat{H} + \beta \widehat{H}_\omega + X)} \cr
c_2 & = \frac{\beta}{2} \frac{ g (\beta \widehat{H} + \beta \widehat{H}_\omega + X) g'' (\beta \widehat{H} + \beta \widehat{H}_\omega + X) - g' (\beta \widehat{H} + \beta \widehat{H}_\omega + X)^2 }{g (\beta \widehat{H} + \beta \widehat{H}_\omega + X)^2}
\end{align}
where the primes denote derivatives.

Similarly for the entropy production we have
\be \label{EPexp}
\sigma_V = - \delta_{\widehat{\Psi}} \left( \sum_{p=0}^\infty \lambda^p \, c_p \, V^p \right) = \sum_{p=0}^\infty \lambda^p \, c_p \left( - \delta_{\widehat{\Psi}} (V^p) \right) = \sum_{p=0}^\infty \lambda^p \, \sigma_V^{(p)} \, ,
\ee
since the functions $c_p$ only depend on $(\widehat{H} , \widehat{H}_\omega , X , \beta )$. Consequently we can write the fluxes as
\be
\mathbf{\Theta}_k = \sum_{p=0}^\infty \lambda^p \, \mathbf{\Theta}_k^{(p)} \, .
\ee
On the other hand we know from \eqref{NESSru} that for any NESS $\widehat{\chi}^+ \in \Sigma^+_V \left( \widehat{\Psi} \right)$ we can expand
\be
\chi^+ (\sfa) = \sum_{s=0}^\infty \lambda^s \, \chi^+_s (\sfa)
\ee
for any $\sfa \in \cA^{(b)}_{R,V}$. Therefore entropy production $\mathrm{Ep} (\widehat{\chi}^+)$ is in principle computable as a power series in $\lambda$.

\subsection{A toy model example}

To summarize and illustrate some physical consequences of the formalism, we present a simple toy model. We couple the system to a single bath and focus on the interacting right algebra $\cA_{R,V}^{(b)}$ discussed in \ref{IntAlgandKMS}. This coupling allows for energy exchange with the bath, giving rise to interesting thermodynamic effects.

To capture the basic idea, we can model the process where Hawking radiation enters the bath with an unrealistic but simple interaction term that destroys quanta in the bulk and creates them in the bath, and viceversa:
\be \label{toyV}
V_R = \lambda \, \left( \sfa \otimes \mathsf{d}_k^\dagger + \sfa^\dagger \otimes \mathsf{d}_k \right) \, .
\ee
In this simplified model, there is a single Hawking particle in the bulk, represented by the operator $\sfa$. We suppress additional labels for simplicity. The bath operator $\mathsf{d}^\dagger_k$ creates quanta in the bath in the state $k$ of the bath’s Fock space. The process occurs with a fixed rate proportional to $\lambda^2$. While this interaction is too simplistic to model the actual exchange of Hawking radiation between the bulk and the bath (see for example \cite{Almheiri:2019psf, Almheiri:2019qdq, Penington:2019npb}), it serves to illustrate the formalism. After taking the crossed product the relevant interaction term becomes \eqref{cVint}, which contains \eqref{toyV} but also logarithmic terms.

For the bath, we take the simple Hamiltonian $H = \sum_j \varepsilon_j \mathsf{d}^\dagger_j \mathsf{d}_j$. Since we are only considering the right algebra, we suppress the $R$ and $L$ labels. We can use this setup to compute some of the thermodynamic quantities we have discussed in this Section. For example, let us consider the energy flux \eqref{ThetaFlux} associated with the right algebra:
\be
\Theta = \ii \left[ \sum_j \varepsilon_j \mathsf{d}^\dagger_j \mathsf{d}_j \, , \, \cV \right] \, .
\ee
We use the expansion \eqref{expV} to write at first order:
\be
\cV = c_0 + \lambda V \, c_1 + \cO (\lambda^2) \, .
\ee
Note that higher-order terms arise due to the nonlocal structure of the modular Hamiltonian. This is a gravitational effect, absent in a purely quantum mechanics setup. Since the coefficients $c_0$ and $c_1$ commute with the bath Hamiltonian, the first non trivial term which contributes to a transition amplitude is the commutator:
\begin{align}
\Theta^{(1)} = c_1 \ii \lambda \left[ \sum_j \varepsilon_j \mathsf{d}^\dagger_j \mathsf{d}_j \, , \, \sfa \otimes \mathsf{d}_k^\dagger + \sfa^\dagger \otimes \mathsf{d}_k \right] = c_1 \ii \lambda \, \varepsilon_k \left( \sfa \otimes \mathsf{d}_k^\dagger - \sfa^\dagger \otimes \mathsf{d}_k \right) \, .
\end{align}
This corresponds to the operator that transfers the particle from the bulk to the bath (and vice versa), weighted by the energy $\varepsilon_k$. This operator can therefore be interpreted as the one responsible for the energy (or heat) transfer. Note that the coefficient $c_1$ is another gravitational effect, coming from the nonlocal structure of the modular Hamiltonian. Higher-order terms in $\lambda$ can be computed similarly. Furthermore, one can compute a formal entropy production observable to first order in $\lambda$ by using \eqref{sigmaV2} and the expansion \eqref{EPexp}:
\be
\sigma_V^{(1)} = - \beta \Theta^{(1)} \, ,
\ee
having dropped a total derivative. When computed on a certain state this operator measures the entropy produced in the process. This model is likely too simple to exhibit a nonequilibrium steady state. However, it effectively captures the basic ideas presented in this paper and shows how quantities are explicitly computable.

\section{Some applications} \label{applications}

In this Section we outline two applications of the formalism. Each one should be explicitly computable using the perturbative expansion we have discussed in the previous Sections, once a specific form for the interaction is chosen.

\subsection{Evaporating black hole}

Now we want to discuss how the formalism could be in principle applied to the case of an evaporating black hole. 

So far we have studied gravitational algebras which arise in the background of the eternal black hole, coupled to reservoirs. We would like to argue that the formalism we have introduced can be useful also in the case of an evaporating black hole. 

To begin with we imagine coupling the system only to a single reservoir on the right side, as described in \cite{Almheiri:2019psf}. We assume that the coupling can be at least in part accomplished by a local interaction $V$. For example the interaction can destroy quanta in the bulk and create them in the reservoir.

Ideally we would like to argue that after the black hole has reached thermal equilibrium, we can couple it to the reservoir and observe its evaporation until we arrive at the final pure state. However this process is outside the validity of our approximations. We have always assumed that the bulk physics do not alter significantly the properties of the reservoir. The latter is assumed to be some infinite system at some fixed temperature, in this case lower than the black hole. It is not clear that this assumption holds throughout the black hole evaporation.

We can however follow the process in different stages by appropriately introducing time scales. Consider the black hole in thermal equilibrium. Then we couple it to the reservoir for a fixed interval of time, centered around some time $T_1$. Then our formalism describes accurately the NESS where a steady flux of radiation enters the reservoir, before the interaction is shut down. The $t \longrightarrow \infty$ limit in the definition of the NESS, in this context means that we wait enough time after the interaction is turned on so that a steady state is reached, but not so much that the interaction is shut down. After the interaction is shut down the black hole is again an isolated system and it will thermalize. We can imagine repeating this procedure for a collection of parametrically large times $T_a$ before and after the Page time. Hopefully the study of the NESS around each $T_a$ contains information about the Page curve. We hope to return to this setting in the future. For the moment let us discuss more precisely the setup and the kind of qualitative information given by the entropy production formula.

We can be more precise following a procedure outlined in \cite{Chandrasekaran:2022eqq}. To model the system we begin with a shell of matter which collapses and forms a black hole. After a parametrically large time $T \gg \beta$ the black hole will have reached thermal equilibrium. This means that in the strict large $N$ limit the correlators of single trace operators will look thermal. We can therefore construct a large $N$ algebra $\cA_{0,R}$. If we chose $T$ so that $T \longrightarrow \infty$ when $N \longrightarrow \infty$, then finite time $t \ll T$ evolution will preserve the algebra $\cA_{0,R}$, and in particular will preserve thermal equilibrium.

More precisely we will have an algebra $\cA_{0,R}$ of early time operators $\sfa (t)$ and an algebra $\cB_{0,R}$ of late time operators $\sfb (T + t')$. In the large $N$ limit correlators involving operators $\sfa (t)$ and $\sfb (T + t')$ will factorize into products of early time and of late time operators. In the large $N$ limit finite time evolution will map $\cA_{0,R}$ to itself. Therefore we can talk about the algebra $\cA_{0,R}$ in thermal equilibrium, and by adjoining the operator $U$ as before, it will become a Type $\mathrm{II}_\infty$ algebra $\cA_R$ \cite{Chandrasekaran:2022eqq}. 

Now that we know how to define $\cA_R$, we can couple it to a reservoir as discussed in the paper, for a fixed but parametrically small amount of time, so that a NESS appears before the coupling is shut off.

We can therefore apply our results to this case. We consider a total system made by the black hole, now in thermal equilibrium, and a single reservoir. As before we take the reference state $\widehat{\Psi} $ in factorized form. 

If we denote by $\delta_{\widehat{\tau}}$ the infinitesimal generator of the decoupled evolution, we have $\delta_{\widehat{\tau}}  =  \delta_R + \delta_I$, where $\delta_I = \ii [I , \cdot]$. Note that the decoupled system is KMS with respect to the evolution $\delta_{\widehat{\Psi}}  = \beta_R \, \delta_R + \beta \, \delta_I$. 

Now we imagine putting the two subsystems in contact by a perturbation $V(t)$ which is vanishing outside an interval $[0 , t_1]$. This perturbation changes the boundary condition at infinity from reflecting to transparent for a finite amount of time. We expect the system to reach a NESS where the black hole slowly evaporates until the perturbation is removed and then thermalizes again.

By the entropy production formula \eqref{EntProd} it follows
\be
S \left( \widehat{\Psi} \circ \widehat{\tau}^t_V \vert \widehat{\Psi} \right) =  - \ii \widehat{\Psi} \left( \Gamma^t_V \, \delta_{\widehat{\Psi} \otimes \omega} (\Gamma^{t*}_V) \right) = - \ii \widehat{\Psi} \left( \Gamma^t_V \, \delta_{I} (\Gamma^{t*}_V) \right) - \ii \widehat{\Psi} \left( \Gamma^t_V \, \delta_{\omega} (\Gamma^{t*}_V) \right) \, .
\ee
What is the physical interpretation of the two terms on the right hand side? Let us explain this with a finite dimensional example.

\paragraph{Example.} Assume we have a finite dimensional system initially in thermal equilibrium. We assume its hamiltonian has the form $H (t) = H + V(t)$ where $ V(t)$ is a time dependent perturbation which in particular vanishes outside a certain interval $[0 , t_1]$. The initial state of the system $\eta$ is $(\tau , \beta)$-KMS with respect to the evolution generated by $\delta_\omega =  \beta \, \delta = \ii \beta \left[ H , \, \cdot \right]$. Then as above the entropy production formula gives
\be \label{EvBH-EntProd}
S \left( \eta \circ \tau^T_V \vert \eta \right) =   \ii \beta \, \eta \left( \Gamma^T_V \, \delta (\Gamma^{T*}_V) \right) \, .
\ee
We seek a physical interpretation of the right hand side. Since the system is finite dimensional, the energy at time $t$ can be computed from the Hamiltonian as
\be
\cE (t) = \eta \left( \tau^t_V (H(t)) \right) \, .
\ee
Note that this quantity would be divergent for an infinite dimensional system. 

Since the perturbation is identically zero outside the time interval $[0, t_1]$ we can compute the difference between the initial and final energies
\begin{align} \label{Ex-work}
\cE (t_1) - \cE (0) = \int_0^{t_1} \eta \circ \tau_V^t \left( \delta (V(t)) \right) \, \dd t =  \ii \eta \left( \Gamma^{t_1}_V \, \delta \left( \Gamma_V^{t_1, \dagger} \right) \right) \, .
\end{align}
Therefore the observable on the right hand side has the physical interpretation as the total amount of work done on the system. Note that while for an infinite dimensional system the energy computed via the Hamiltonian would be divergent, energy differences are well defined and captured by the right hand side of \eqref{Ex-work}. 
$\blacksquare$

Now we can interpret \eqref{EvBH-EntProd} as 
\be
S \left( \eta \circ \tau^T_V \vert \eta \right) =  \beta_R W_R + \beta W_S \, .
\ee

Finally we can repeat the argument to allow for several algebras at different time instants $T_a$. We can adjust the time scales so that the algebras of early time operators and the algebras of late time operators capture physics before and after the Page time. In this way we have all the tools to address thermodynamic aspects of the system before and after the Page time. We hope to return to this setting in the future.

\subsection{Quantum chaos}

In this Section we use the formlism we have built to propose a variant of the OTOC as a tool to probe quantum chaos in operator algebras. 

\paragraph{Finite dimensional systems.}

The OTOC is widely recognised as a diagnostic of quantum chaos \cite{Shenker:2013pqa,Maldacena:2015waa}. It is a time and temperature dependent function which is defined via the averaging of a double commutator
\be \label{OTOC-findim}
\cC_{V,W} (t) = \Tr \left( \rho \, \left[ W_t , V \right]^\dagger \, \left[ W_t , V \right] \right) \, .
\ee
To unpack a bit the notation, here $W_t = \e^{\ii H t} \, W \e^{- \ii H t}$ is the operator evolved in time and $\rho = \e^{- \beta H} / Z$ is the thermal density operator. The operators $V , W \in \cB (\cH)$ are usually taken to be Hermitian and with local support.

Physically the growth of the function \eqref{OTOC-findim} measures the spread of quantum information, also known as \textit{information scrambling} \cite{Shenker:2013pqa}. For example we can choose the operators in such a way that $\left[ W , V \right] = 0$, for example by having them have support on space-like separated regions. A non vanishing $\left[ W_t , V \right] $, and therefore a non vanishing \eqref{OTOC-findim} measure how quickly the operator $W$ spreads among an operator basis. In chaotic systems the OTOC is expected to grow until a certain scrambling time is reached, where any subsystem is now maximally entangled, and then remain approximately constant.

In the particular case where the two operators are position and momentum, the semi-classical limit of \eqref{OTOC-findim} has the exponential behavior $\e^{2 \lambda t}$ where $\lambda$ can be interpreted as a Lyapunov exponent, which measures the sensitivity to initial conditions of classical trajectories.

\paragraph{Quantum chaos in gravitational algebras.}

Consider now a quantum dynamical system $\left( \cO , \alpha \right)$ where $\cO$ is a von Neumann algebra and $\alpha$ determines the time evolution. Consider a reference state $\psi$ which we assume to be $(\alpha , \beta)$-KMS. In the general infinite dimensional setting we won't in general have a finite Hamiltonian operator or a Gibbs state, but we might replace them with the $*$-automorphism which governs the time evolution and a general normal state $\psi$. Then given two operators $\sfv , \sfw \in \cO$ we define the \textit{algebraic OTOC} as
\be \label{OTOC-vn}
\sfC_{\sfv , \sfw} (t) = \psi \left( \left[ \alpha^t (\sfw) , \sfv \right]^\dagger \left[ \alpha^t (\sfw) , \sfv \right]  \right) \, .
\ee
This definition is more general than \eqref{OTOC-findim} and reduces to it in the finite dimensional setting. It however holds for an arbitrary normal state $\omega$. See also \cite{Furuya:2023fei,Gesteau:2023rrx,Ouseph:2023juq} for a different approach.

Oddly such definition appears to be new in the context of the operator algebra approach to quantum statistical mechanics. An even more natural definition appears in the case where $\psi$ is modular, where we can replace the $*$-automorphism $\alpha$ with the modular flow associated to $\psi$. In this case we can study the OTOC using the structural properties of the modular flow. Denote by $\Delta_\psi$ the modular operator associated to $\psi$. In this case we define a \textit{modular} algebraic OTOC as
\be
\scrC_{\sfa , \sfb} (s) = \psi \left( \left[ \Delta_\psi^{- \ii s} \sfa  \Delta_\psi^{\ii s} ,\sfb \right]^\dagger \left[\Delta_\psi^{- \ii s} \sfa  \Delta_\psi^{\ii s} ,\sfb \right]  \right) \, .
\ee
We could be even more general and replace the state $\psi$ with an arbitary state $\chi$, for example a NESS.

Now we turn our attention to the gravitational algebras in black hole backgrounds. In this case the relevant operator algebra is the algebra $\cA^{(b)}_{R,V} $ discussed in Section \ref{IntAlgandKMS}. In this case we have the modular vector $\widehat{\Psi}_V$ and the modular-time evolution operator $\widehat{\Delta}_{\Psi_V}$. This  induces the inner $*$-automorphism
$ \widehat{\tau}^t_V  $ introduced in \eqref{tauVhat}.

Now for any pair of operators $\widehat{\sfa}_V , \widehat{\sfb}_V \in \cA^{(b)}_{R,V} $ we can write
\be \label{OTOC-gr}
\scrC_{\sfa , \sfb} (s) = \chi \left( \left[ \tau_V^s (\widehat{\sfa}_V) , \widehat{\sfb}_V \right]^\dagger \left[ \tau_V^s (\widehat{\sfa}_V) , \widehat{\sfb}_V \right]  \right) \, .
\ee
In this case the state $\chi$ could be a non-trivial NESS or an equilibrium KMS state. Let us unpack a bit this definition. For simplicity let us assume that the two operators $\widehat{\sfa}_V , \widehat{\sfb}_V \in \widehat{\cA}_V$ are Hermitian, so that we can rewrite \eqref{OTOC-gr} as
\be
\scrC_{\sfa , \sfb} (s) = - \chi \left( \left[ \tau_V^s (\widehat{\sfa}_V) , \widehat{\sfb}_V \right]^2  \right) \, .
\ee
To be concrete we can choose to be in thermal equilibrium, where we already expect to see some signatures of chaos and pick as $\chi$ the perturbed KMS state $\widehat{\Psi}_V$
\be
\scrC_{\sfa , \sfb} (s) = - \braket{ \widehat{\Psi}_V \vert \left[ \tau_V^s (\widehat{\sfa}_V) , \widehat{\sfb}_V \right]^2 \vert \widehat{\Psi}_V }
\ee
We propose this function as a probe of quantum chaos in operator algebras when gravitational effects are included. Since $\cN=4$ SYM above the Hawking-Page transition is expected to be maximally chaotic, we expect that these functions should have very interesting properties.

Note that the modular OTOC could be as well studied in other systems, including the microcanonical gravitational algebra introduced in \cite{Chandrasekaran:2022eqq}.

\section{Conclusions}

In this note we have studied certain aspects of the gravitational algebras introduced in \cite{Witten:2021unn} when driven out of equilibrium. Our primary technical tool has been to couple the original system with an external reservoir.  Once this coupling is established, it becomes conceivable to manipulate it, thereby driving the original system out of equilibrium and allowing for the exploration of its thermodynamic properties, including entropy production. In order to do so we have found an explicit expression for the quantum dynamics of the system and we have use it to set up a perturbative treatment for the nonequilibrium steady states and for the relevant observables. In principle everything is computable order by order. From the technical point of view, these results were obtained by generalizing several findings from nonequilibrium statistical mechanics, often written with the type $\mathrm{I}$ algebra of a finite system in mind, to encompass type $\mathrm{II}_\infty$ gravitational algebras. 

We have however merely scratched the surface of the subject as many interesting open problems remain. We would like to highlight some of these challenges:
\begin{itemize}
\item The primary limitation of this note lies in the formal nature of the construction. One can only go so far without specifying a detailed interaction term and making other choices. To make progress it would be necessary to choose a certain setting and work out explicit examples. A natural place to start is the case of JT gravity where the algebraic formalism has been established in detail in \cite{Penington:2023dql}.
\item Ideally the main goal of this program would be to say something about the Page curve. The evaporation process of a black hole inherently involves nonequilibrium dynamics. Hopefully the relation between entropy and nonequilibrium dynamics we have discussed could be useful to connect the results of \cite{Almheiri:2019psf,Penington:2019npb} with the algebraic formalism.
\item While we have discussed several aspects of nonequilibrium physics, we have yet to introduce two fundamental tools of nonequilibrium quantum statistical mechanics: the spectral analysis and the scattering approach \cite{JP1,RuelleRev}. Both tools require appropriate generalization to be applied in our context.
\item It should be possible to generalize our discussion to the microcanonical ensemble introduced in \cite{Chandrasekaran:2022eqq}. One can for example imagine to perturb the boundary theory with certain operators and appropriately tuning the couplings to drive the system out of equilibrium. 
\item In Section \ref{NEdyn} we have chosen the simplest way to implement the $1 / N$ corrections via the crossed product. There could be more interesting ways which involve the reservoirs in an essential manner. For example this could provide an avenue to include gravitational effects in the reservoirs.
\item While the OTOC is widely recognized as a probe of quantum chaos, there does not appear to be any comprehensive discussion in the context of Tomita-Takesaki theory, although see also \cite{Furuya:2023fei,Gesteau:2023rrx,Ouseph:2023juq}. Possibly such a link should be established before attempting to use the OTOC in the context of gravitational algebras. It is fascinating to speculate that the chaotic behavior of $\cN=4$ SYM above the Hawking-Page transition is connected with the appearance of the $\rm{II}_\infty$ factor. It would also be interesting to see if this holds in simpler systems in quantum statistical mechanics.
\item It has been recently shown in \cite{AliAhmad:2023etg,Klinger:2023tgi,Kudler-Flam:2023hkl} that the crossed product construction can be applied to quantum field theories beyond the original holography framework. Therefore the results presented in this note could in principle offer a new perspective on nonequilibrium dynamics in generic quantum field theories. Investigating this further would be a very interesting avenue of research.
\item In this note, we have not considered symmetries. As in \cite{Witten:2021unn}, the crossed product construction involved only the modular automorphism group. However, \cite{AliAhmad:2024eun} identifies conditions under which the crossed product yields a semifinite algebra for more general symmetry groups. The crossed product on backgrounds with symmetries is also discussed in \cite{Fewster:2024pur}. This perspective may lead to a generalization of the formalism presented here to study nonequilibrium dynamics in systems with charges. This raises interesting questions, such as under what conditions symmetries are broken by nonequilibrium dynamics or whether new symmetries could emerge in nonequilibrium states.
\item Finally one of the primary motivations behind this note was to derive a gravitational analog of the many nonequilibrium thermodynamics identities, such as Jarzynski's or Crook's work relations. Some progress in this direction is reported in \cite{Cirafici:2024ccs}.
\end{itemize}
We hope to report on progress about these challenges in the near future.

\vskip0.5cm
 \noindent {\bf {Acknowledgements.}} 
 I am supported by INFN via the Iniziativa Specifica GAST and I am a member of IGAP and of INDAM-GNFM.
 \bibliographystyle{unsrt}

\end{document}